\documentclass[aps,pra,reprint,showpacs,longbibliography]{revtex4-1}
\usepackage{amsmath,amssymb,graphicx,units}
\usepackage[usenames,dvipsnames]{color}

\begin{document}
\title{Many-body delocalization with random vector potentials}
\author{Chen Cheng}
\author{Rubem Mondaini}
\affiliation{Beijing Computational Science Research Center, Beijing 100193, China}

\begin{abstract}
We study the ergodic properties of excited states in a model of interacting fermions in quasi-one-dimensional chains subjected to a random vector potential. In the noninteracting limit, we show that arbitrarily small values of this complex off-diagonal disorder trigger localization for the whole spectrum; the divergence of the localization length in the single-particle basis is characterized by a critical exponent $\nu$ which depends on the energy density being investigated. When short-range interactions are included, the localization is lost, and the system is ergodic regardless of the magnitude of disorder in finite chains. Our numerical results suggest a delocalization scheme for arbitrary small values of interactions. This finding indicates that the standard scenario of the many-body localization cannot be obtained in a model with random gauge fields.
\end{abstract}

\maketitle

\section{Introduction}

The topic of localization in disordered quantum systems is long-standing~\cite{Anderson1958,Evers_Mirlin_2008,Abrahams1979}. While Anderson~\cite{Anderson1958} initially
showed that disorder on on-site energy levels drives localization of single-particle states, it took about 50 years to have a better understanding of how interactions affect this picture.. For small values of interactions, perturbative calculations~\cite{Fleishman80,Altshuler_Gefen_97,Gornyi_Mirlin_05,Basko2006,Basko2008} have demonstrated that under certain conditions localization at high temperatures still occurs. This has been also confirmed via several numerical simulations of different quantum models~\cite{Oganesyan_Huse_07,Znidaric08,Pal10,Khatami_Rigol_12,Bardarson_Pollmann_12,Luitz2015,Lev2015a,Mondaini2015}. Not only of theoretical relevance, this phenomenon, dubbed many-body localization (MBL)~\cite{Nandkishore_Huse_review_15,Altman_Vosk_review_15}, has recently been explored experimentally with trapped ultracold atoms~\cite{Schreiber2015,Bordia2016,Choi2016} and ion chains~\cite{Smith2015}.

One of the remarkable consequences of MBL is its manifest breakdown of the scheme proposed by the eigenstate thermalization hypothesis (ETH)~\cite{Deutsch1991,Srednicki1994,Rigol2008,Dalessio2016}, which explains how a generic isolated quantum system thermalizes when taken far from equilibrium without the necessity of coupling it to an external bath. In the presence of sufficiently strong disorder, however, thermalization induced by interactions is halted, and the quantum system may preserve the memory of its initial conditions for arbitrarily long times. This latter property was recently used as verification of MBL in systems of ultracold atoms in one-~\cite{Schreiber2015} and two-dimensional~\cite{Bordia2016,Choi2016} optical lattices. If properly isolated, MBL systems may also be used as prototypical quantum memories with potential use for information processing in quantum computers~\cite{Serbym2014a,Choi2015,Yao2015,Vasseur2015}.

Most of the studies that investigate the breakdown of thermalization and ergodicity share in common the situation where disorder is introduced by random scalar potentials (diagonal disorder) in analogy with the original Anderson model for localization. Even in the noninteracting limit, further investigations have shown that Anderson-like localization also takes place with real off-diagonal disorder away from singular energies in which the density of states diverges, as in fermionic systems with hopping disorder~\cite{Shapir1982,Altland2001,Xiong2007}. In this paper we address the question of whether a \textit{complex} off-diagonal disorder may display localization in the presence of interactions. Physically, we aim to elucidate if a random vector potential may also result in MBL in quasi-one-dimensional chains. As the random vector potential breaks time-reversal symmetry, it also allows the study of ergodicity (and its potential breakdown) in systems whose corresponding ensemble of random matrices, with similar spectral statistics of the system's Hamiltonian, is the Gaussian unitary ensemble (GUE) rather than the Gaussian orthogonal ensemble (GOE)~\cite{Batsch1996,Geraedts2016}, prominently used in the case of random scalar potentials.

From the experimental point of view, there has been an intense effort to produce artificial gauge fields in optical lattices using different techniques~\cite{Dalibard2011,Goldman2014,Goldman2016}. While some of these methods may inherently generate artificial \textit{homogeneous} gauge fields, for example, by periodically shaking the lattice~\cite{Jotzu2014}, others, such as the ones that explore the synthetic dimensions of the atoms via their internal states with an additional resonant laser field~\cite{Celi2014}, have the potential of being generalized to the situation of site-dependent or even random fields, ultimately allowing the emulation of the model studied in this paper.

\section{Model and localization in noninteracting systems}
\begin{figure}[!tb] %%% FIG1
 \includegraphics[width=0.99\columnwidth]{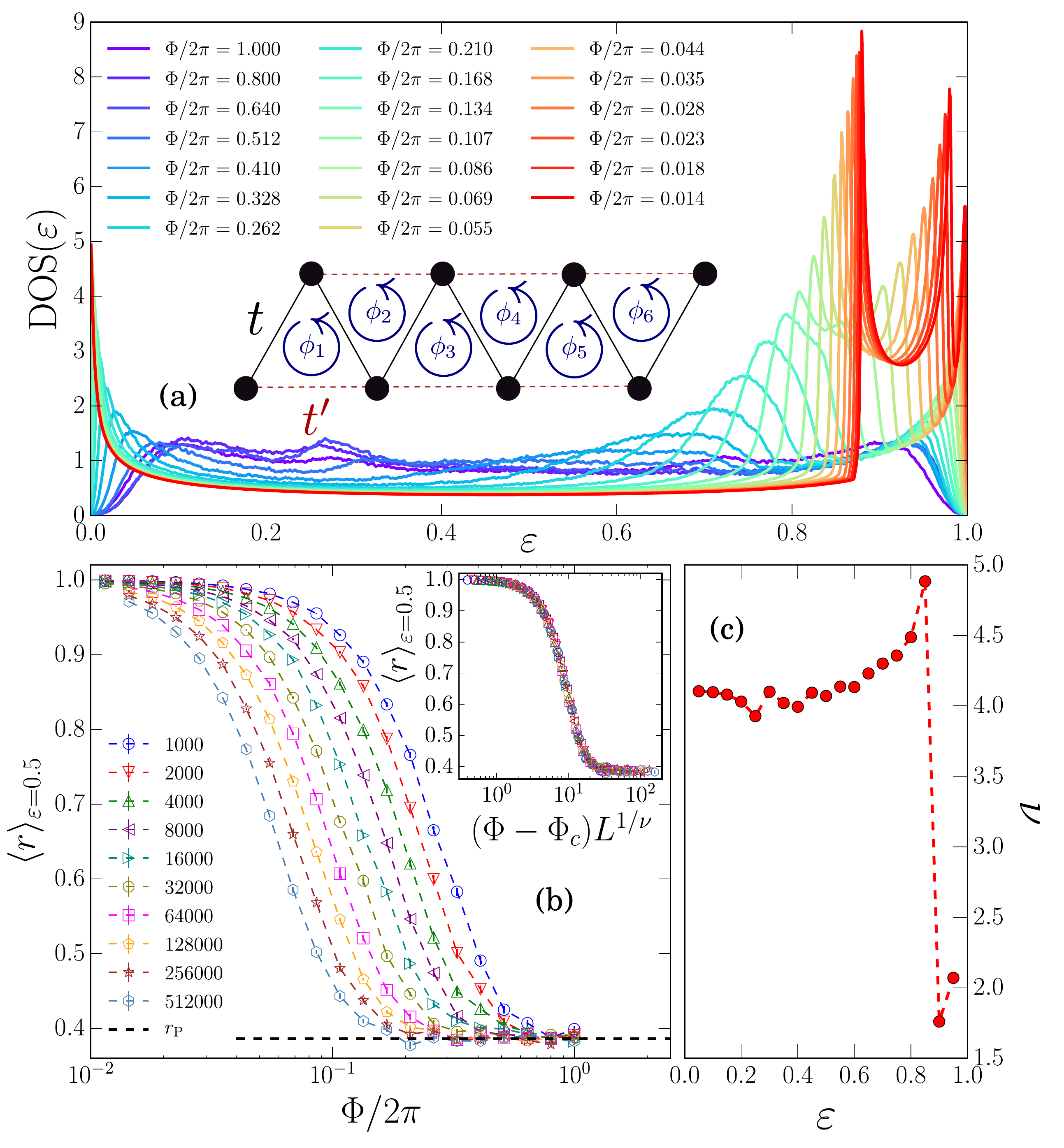}
 \vspace{-0.6cm}
 \caption{(Color online) Results for the noninteracting model. (a) Energy density dependence of the density of states in the single-particle basis for a lattice with $64\ 000$ sites and different disorder amplitudes. (b) The ratio of adjacent gaps for lattice sizes up to $L=512\ 000$ in the middle part of the spectrum and (c) the values of the critical exponent related to the divergence of the correlation length in different parts of the spectrum. We use the method described in Ref.~\cite{Polizzi2009} for the three largest system sizes and full ED for the others. }
 \label{fig:Fig_1_tprime05}
\end{figure}

The model with random vector potential (also known as the random flux model) has one particularity: By performing a suitable local gauge transformation one can eliminate the inhomogeneous Peierls phases in the hoppings, stemming from the vector potential, in one of the spatial dimensions. Hence, the problem is trivial in one-dimensional systems with nearest-neighbor hoppings. Nevertheless, by introducing next-nearest-neighbor hoppings the lattice becomes equivalent to a triangular ladder, in which there is a finite and random magnetic flux for each triangular plaquette. This is the scenario we explore here [see cartoon in Fig.~\ref{fig:Fig_1_tprime05}(a)]. We initially study the localization properties in the following noninteracting Hamiltonian $\hat H_0$:
\begin{equation}
 \hat H_0 = -t\sum_{j=1}^{L-1}
  ( \hat c^{\dagger}_{j} \hat c^{}_{j+1}
  + \text{H.c.}) \nonumber \\
  -t^\prime\sum_{j=1}^{L-2}
  (e^{\imath\phi_j}\hat c_{j}^\dagger \hat c_{j+2}^{}
  + \text{H.c.}),
\label{eq:nonintHamiltonian}
\end{equation}
where $\hat c^{\dagger}_{j}$ ($\hat c_{j}$) creates (annihilates) a spinless fermion on site $j$; $t$ and $t^\prime$ are the hopping amplitudes between the nearest and next-nearest neighbors. The random phases $\phi_j$ are chosen from a symmetric and uniform distribution $\left[-\frac{\Phi}{2},\frac{\Phi}{2}\right]$, and $\Phi$, the disorder amplitude, assumes values in $\left[0,2\pi\right]$. In direct contrast to the random scalar potential, the disorder promoted by the random vector potential is bounded by its maximum value $\Phi_{\rm max}=2\pi$. We restrict ourselves to the situation where $t^\prime=t/2$, with $t=1.0$ setting the energy scale, and use open boundary conditions \footnote{This is to avoid the situation of being close to a point where momenta is a good quantum number for small disorder amplitudes in the case of periodic boundary conditions. That hinders the level repulsion characteristic of ergodic phases~\cite{Oganesyan_Huse_07}} in lattices with $L$ sites. We use full exact diagonalization (ED) for the smaller system sizes while in the larger lattices employing a numerically efficient contour integration technique~\cite{Polizzi2009} to obtain  approximately 600 eigenpairs in the target energy ranges for each disorder realization. It is worth mentioning that $\hat H_0$ breaks time-reversal, particle-hole, and sublattice symmetries, precluding the manifestation of any topological order~\cite{Schnyder2008}.

It is not immediately clear whether the random vector potential induces single-particle localization. The density of states (DOS), displayed as a function of the energy density $\varepsilon \equiv (E-E_0)/(E_{\rm max}-E_0)$ in Fig.~\ref{fig:Fig_1_tprime05}(a), possesses van Hove singularities which are smeared out by increasing disorder. In the low-disorder regime the structure in DOS potentially signals a markedly different behavior for the various energy densities. To track whether localization takes place we calculate the ratio of adjacent gaps $r_n=\min[\delta^E_{n+1},\delta^E_{n}]/\max[\delta^E_{n+1},\delta^E_{n}]$, with the gaps $\delta^E_n \equiv E_n - E_{n-1}$, and $\{E_n\}$ is the sorted list of energy levels~\cite{Oganesyan_Huse_07}. This quantity has been widely used as a probe of chaos in quantum systems~\cite{Iyer2013,Luitz2015,Baoming2015,Mondaini2015,Serbym2016}. In the localized phase, the repulsion of energy levels is lost, and the level spacings display a Poisson distribution with $r_{\rm P} =2\ln2-1\approx0.386$~\cite{Atas2013}.  Figure~\ref{fig:Fig_1_tprime05}(b) shows the disorder-averaged ratio of adjacent gaps $\langle r\rangle$ (throughout the paper $\langle\cdot\rangle$ denotes disorder averaging) at the energy density $\varepsilon = 0.5$ as a function of the disorder amplitude. Here and in the following, for the case of full ED we get $\langle r\rangle_\varepsilon$ by averaging the results over a window $\Delta\varepsilon=0.02$ around the targeted energy density $\varepsilon$. All the different lattice sizes eventually reach the Poisson limit for a given $\Phi_c(L)$, indicating localization. To obtain the critical disorder $\Phi_c$ in the thermodynamic limit we use a function of the form $f\left[(\Phi-\Phi_c)L^{1/\nu}\right]$~\cite{Luitz2015} which scales the ratio of adjacent gaps [see inset in Fig.~\ref{fig:Fig_1_tprime05}(b)]. Remarkably, $\Phi_c$ is zero for the whole range of energy densities, which is similar to the situation of random scalar potentials in one-dimensional lattices. The critical exponent $\nu$, which is related to the divergence of the single-particle correlation length near a critical energy density $\varepsilon_c$ by $\xi (\varepsilon)\sim|\varepsilon-\varepsilon_c|^{-\nu}$, possesses the energy dependence displayed in Fig.~\ref{fig:Fig_1_tprime05}(c). It is not possible, however, to rule out the scenario of having an exponentially small fraction of the states with a diverging single-particle localization length. This was proposed in Ref.~\cite{Nandkishore2014}, and systems described by the Hamiltonian $\hat H_0$  may then display a \textit{marginal} Anderson localization~\cite{Altland1999}.

\begin{figure}[ht]
  \centering
  \includegraphics[width=0.99\columnwidth]{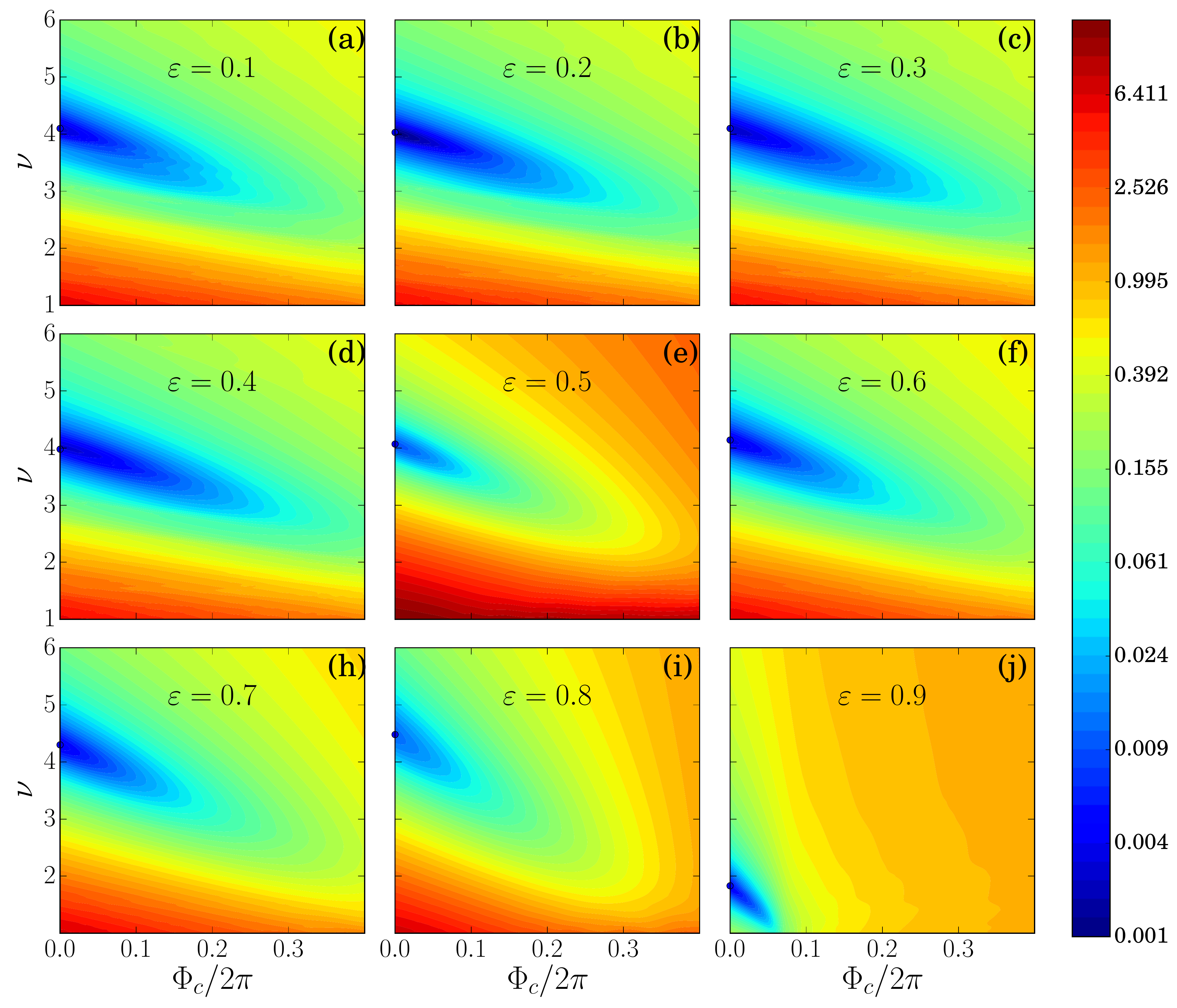}
  \caption{ (Color online) Colored contour plot of the fitting error as a function of  $\Phi_c$ and $\nu$. Here the polynomial degree used in the fittings is $n=12$.}
  \label{fig:fitting_error}
\end{figure}

Some details of the scaling procedure are in order. To carefully obtain the critical disorder $\Phi_c$ in the thermodynamic limit, we fit polynomials of degree $n$ to the data of $r$ and $(\Phi-\Phi_c)L^{1/\nu}$ by using the least-squares method. In Fig.~\ref{fig:fitting_error} we plot the corresponding fitting error for $n=12$ in the two-dimensional parameter space \{$\Phi_c$, $\nu$\} at different energy densities. Essentially, this method allows us to simultaneously estimate the best set of parameters that scale the quantity under investigation. The best values of  $\Phi_c(\varepsilon,n)$ and $\nu(\varepsilon,n)$ are determined by the points with the smallest fitting errors, marked by the dark blue dots in the plots (see Appendix A for an analysis of the polynomial order used in the extraction of best values).

\begin{figure}[ht]
  \centering
  \includegraphics[width=0.99\columnwidth]{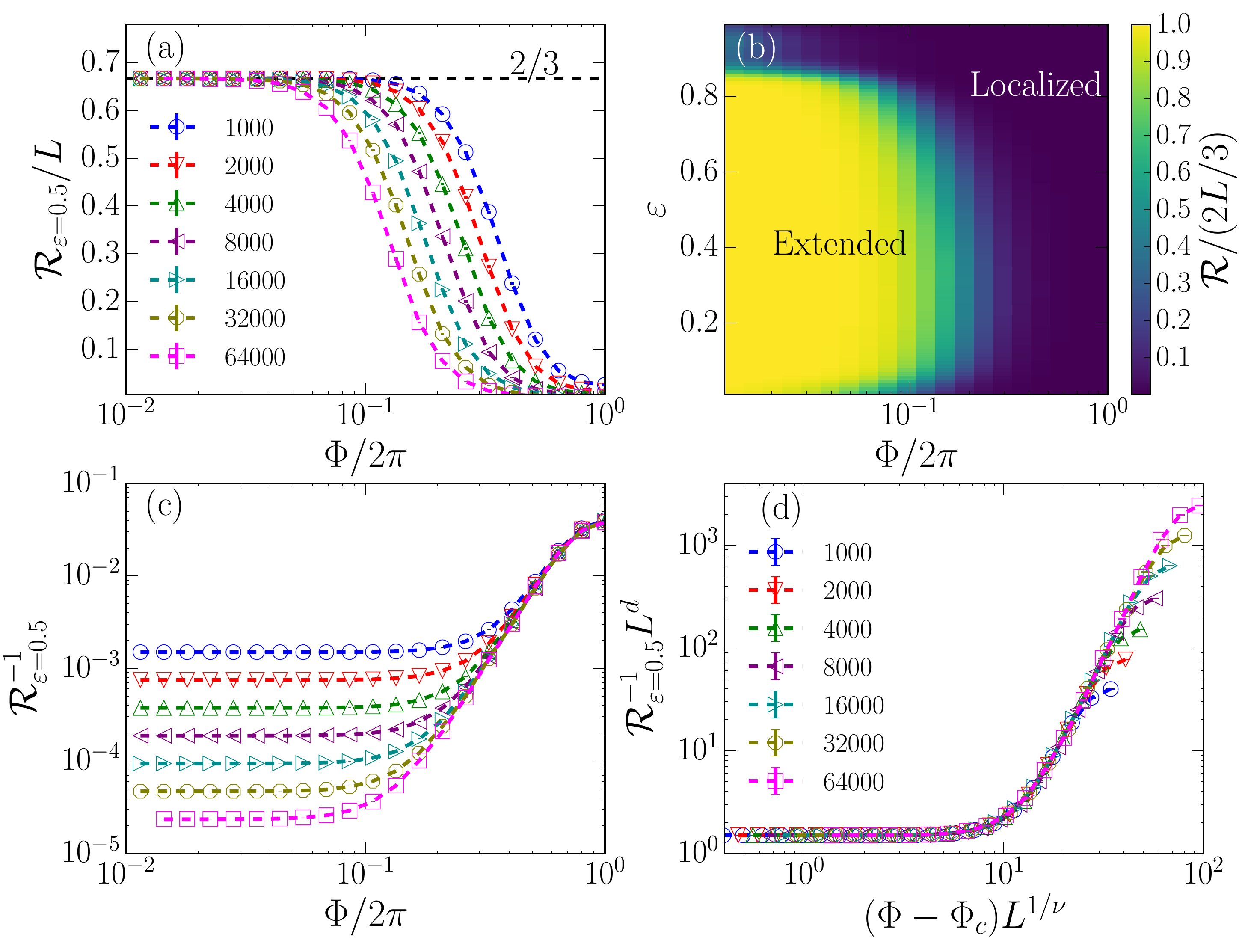}
  \caption{ (Color online) (a) ${\cal R}/L$ and (c) ${\cal R}^{-1}$ for different system sizes up to $L=64\ 000$ in the middle part of the spectrum ($\varepsilon = 0.5$). (b) ${\cal R}/(2L/3)$ as a function of $\Phi/2\pi$ and $\varepsilon$ for $L=16\ 000$. (d) Data collapse of the finite-size scaling.}
  \label{fig:IPR}
\end{figure}

Apart from the ratio of adjacent gaps which examine the degree of ergodicity of the system, manifest in the eigenvalue properties, we also investigate the degree of delocalization of the eigenstates. For the $\alpha$th single-particle eigenstate $|\alpha\rangle=\sum_j\phi^{(\alpha)}_j|j\rangle$, where $|j\rangle$ denotes the real-space single-particle basis state and $\phi^{(\alpha)}_j$ is its corresponding weight in the eigenstate, one can calculate the inverse participation ratio (IPR):
${\cal R}^{(\alpha)}=1/(\sum_j |\phi^{(\alpha)}_j|^4)$. In the case of extended eigenstates, the amplitudes $\phi_j$ are independent random variables resulting in an IPR proportional to the system size $L$. In contrast, for the case of localized eigenstates, it is a constant value independent of $L$. Thus, the normalized value ${\cal R}\times L^{-d}$ ($d$ is the real space dimensionality of the system, i.e., $d=1$ in our case) can be used as an ``order parameter'' detecting the delocalization-localization transition. As shown in Fig.~\ref{fig:IPR}(a), for all system sizes, ${\cal R}_{\varepsilon=0.5}/L$ is equal to 2/3 when the disorder $\Phi$ is weak.
In turn, as we increase $\Phi$, the value of ${\cal R}_{\varepsilon=0.5}/L$ drops quickly to a small value proportional to $1/L$ [see merging of the data in Fig.~\ref{fig:IPR}(c)]. In agreement with the analysis of the ratio of adjacent gaps, this suggests a mechanism of induced localization after increasing disorder. To highlight the importance of a proper finite-size scaling analysis, in Fig.~\ref{fig:IPR}(b) we show a finite-size phase diagram in the parameter space $\{\varepsilon,\Phi/2\pi \}$ for a system with $L=16\ 000$. From this picture we can infer the transition between extended states and localized states as being associated with whether the IPR is extensive or not. From that one would wrongly conclude that the disorder strength that leads to localization is finite and also is smaller for larger energy densities.

However, similar to the ratio of adjacent gaps $r$, the IPR can also satisfy a finite-size scaling form near the transition point as ${\cal R}^{-1}\times L^{d} \sim f\left[(\Phi-\Phi_c)L^{1/\nu}\right]$. The obtained values of $\Phi_c$ and $\nu$ after the scaling agree with the ones obtained from the scaling of $r$; the corresponding data collapse for energy density $\varepsilon = 0.5$ is shown in Fig.~\ref{fig:IPR}(d).

\section{Ergodicity in interacting Hamiltonians}
\begin{figure}[!t] %%% FIG2
 \includegraphics[width=0.99\columnwidth]{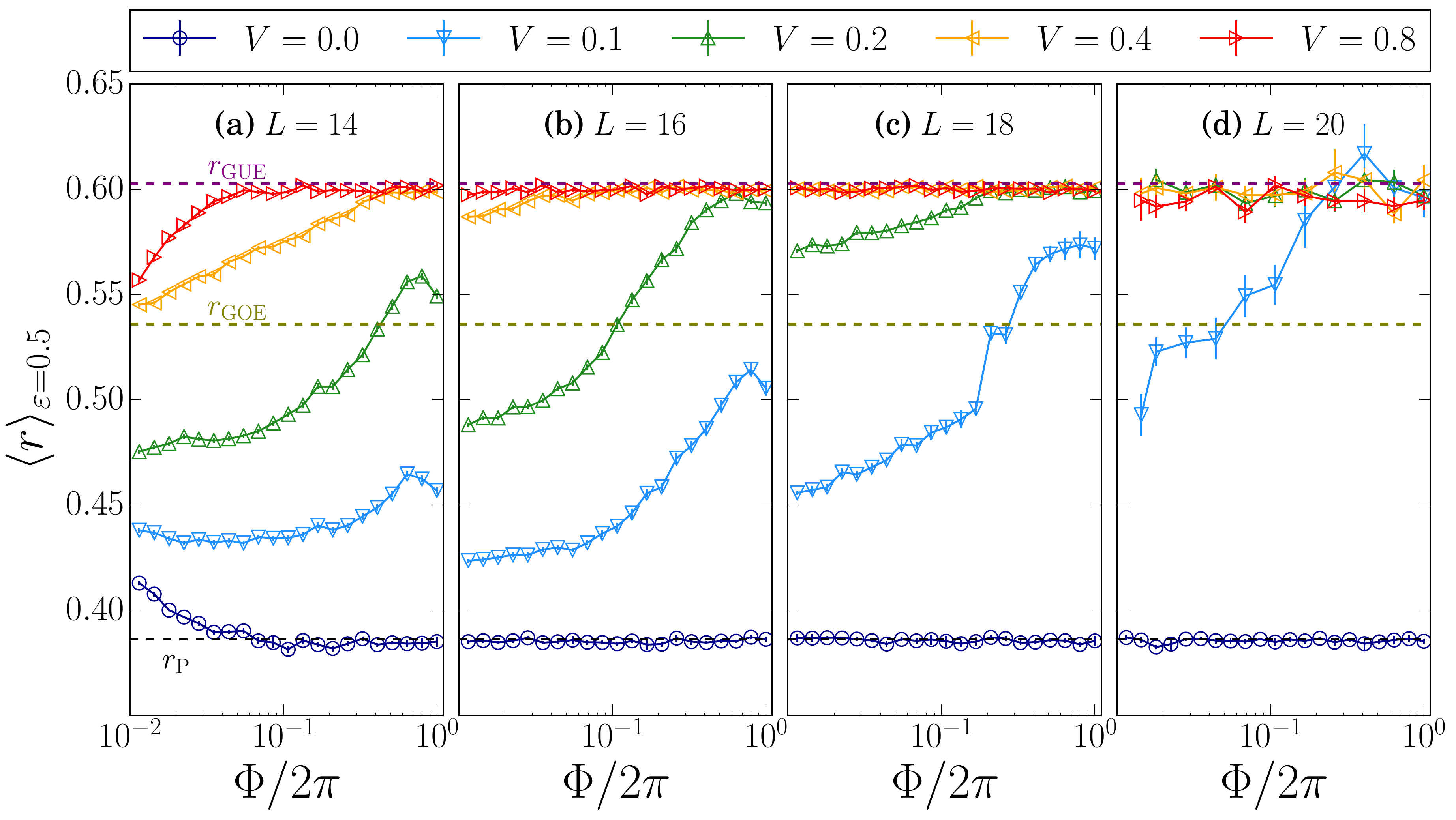}
 \vspace{-0.5cm}
 \caption{(Color online) Disorder-averaged ratio of adjacent gaps at energy density $\varepsilon=0.5$ for increasing NN interactions $V$ and different system sizes, $L=14,16,18$, and 20 in (a)-(d), respectively. For the largest system size we employ the \textsc{FEAST} numerical library~\cite{Polizzi2009} for obtaining approximately 50 eigenvalues at the corresponding energy range.
 }
 \label{fig:roag_interacting_tprime05}
\end{figure}

Motivated by the single-particle localization in the noninteracting case, we now investigate whether a random magnetic field can also promote localization in interacting Hamiltonians $\hat H = \hat H_0 + \hat H_I$. We introduce the nearest-neighbor interactions between the spinless fermions ($\hat H_I=V\sum_{i}\hat n_i \hat n_{i+1}$) with strength $V$ ($\hat n_i \equiv \hat  c^\dagger_i \hat c_i$ is the fermion number operator) and restrict the study to half filling \footnote{Typical equilibration times are defined as when the smallest time in which results of the relaxation dynamics remains closer to the DE prediction.}. Again, to analyze the localization-ergodicity properties, now in the many-body basis, we display in Fig.~\ref{fig:roag_interacting_tprime05} the ratio of adjacent gaps as a function of the disorder amplitude $\Phi$ for increasing values of interaction and different system sizes at the middle of the spectrum ($\varepsilon=0.5$). In the noninteracting case ($V=0$), $\langle r\rangle$ fluctuates around the Poisson limit, suggesting localization. On the other hand, for finite interactions, the ratio of adjacent gaps departs from that limit, and when $V=0.8$, it reaches the GUE limit ($ r_{\rm GUE}\approx0.60266$ ~\cite{Atas2013}) for the whole range of disorder amplitudes studied, even for the smaller lattice $L=14$, indicating ergodic behavior. As expected, finite-size effects are more prominent at small values of $V$, but the tendency to delocalization is clear in the thermodynamic limit regardless of the disorder strength. This can be observed by comparing the values for different system sizes [Figs.~\ref{fig:roag_interacting_tprime05}(a)-\ref{fig:roag_interacting_tprime05}(e)] and noting that $\langle r\rangle$ monotonically tends to $r_{\rm GUE}$ when $L$ increases. Thus, in contrast to the random scalar potentials where localization is eventually recovered at a critical value of the disorder amplitude, within the random vector potential scheme, delocalization promoted by interactions is robust regardless of the disorder strength.

This trend is not particular to the middle part of the spectrum. Figure \ref{fig:r_vs_e} displays the disorder-averaged ratio of adjacent gaps for different energy densities (see Appendix C for the corresponding distributions). First, for the smallest system size, $L=14$, the strong fluctuations for weaker disorders is largely suppressed by increasing the disorder amplitude. This is prominent not only in the noninteracting limit [Fig.~\ref{fig:r_vs_e}(a)] but also for finite $V$'s [Figs.~\ref{fig:r_vs_e}(d) and \ref{fig:r_vs_e}(g)]. In this sense one can notice that the random vector potential aids in reducing the finite-size effects. Second, in the interacting situation [see, for example, Figs.~\ref{fig:r_vs_e}(d)-\ref{fig:r_vs_e}(f) for $V=0.1$], one can observe that the increasing disorder promotes a general trend to depart from values closer to the Poisson limit towards the GUE average value. For larger interactions, $V=0.4$ [Figs.~\ref{fig:r_vs_e}(g)-\ref{fig:r_vs_e}(i)], the ratio of adjacent gaps already reaches the ergodic value $\langle r\rangle = r_{\rm GUE}$, and the range in energy densities in which this is true becomes wider as one increases the system size. This is indicative that in the thermodynamic limit one would not observe a many-body mobility edge.
\begin{figure}[!t]
  \centering
  \includegraphics[width=0.99\columnwidth]{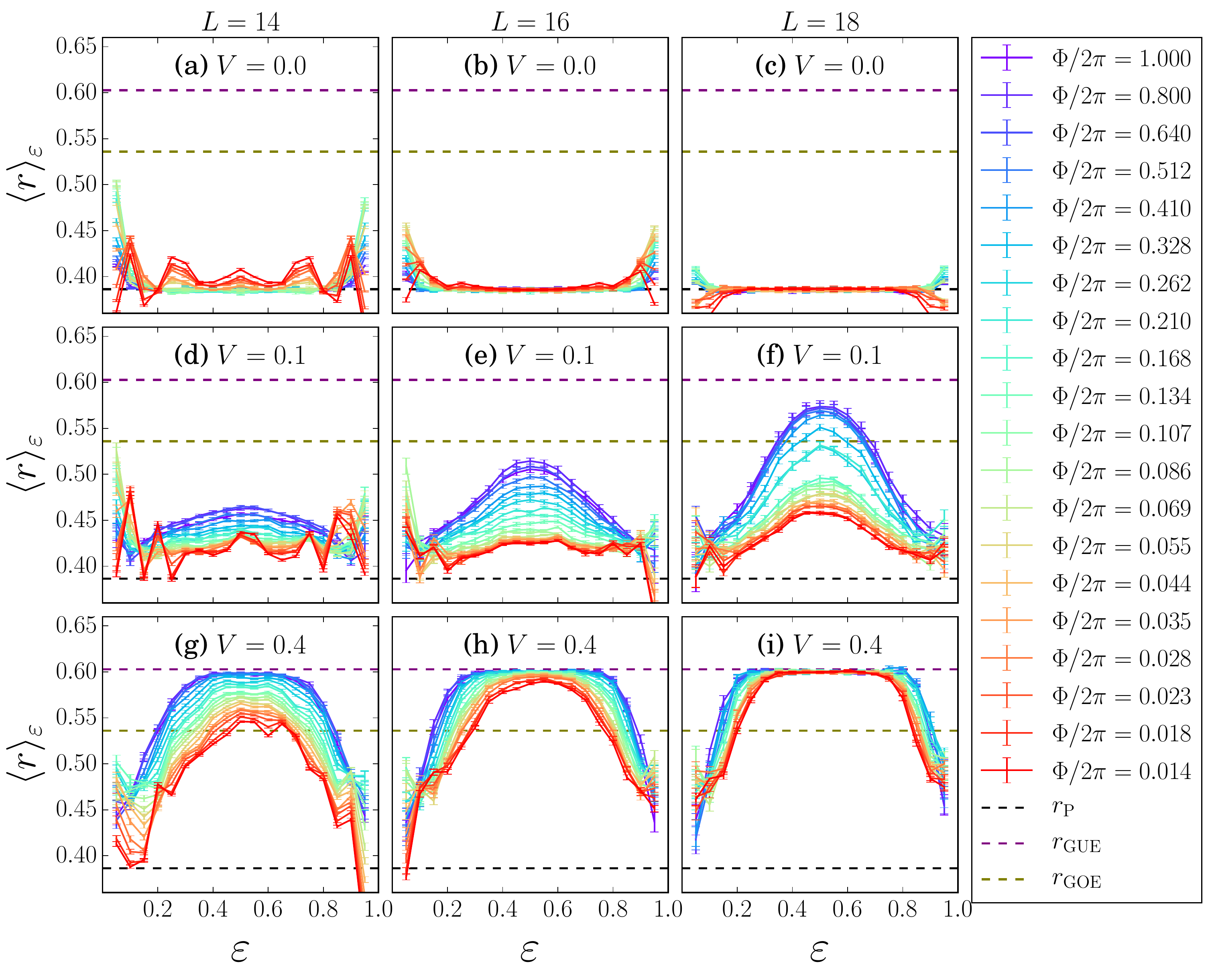}
  \caption{ Disorder-averaged ratio of adjacent gaps as a function of energy density $\varepsilon$ for system sizes $L=14, 16$, and $18$ and interactions $V = 0.0, 0.1$, and $0.4$ with different disorder strengths.}
  \label{fig:r_vs_e}
\end{figure}

\section{Quench Dynamics}
The analysis of the ergodicity of the spectrum, while useful from the theoretical point of view to probe the transition between thermalization and MBL, is, in general, elusive in experiments. Only in very particular setups~\cite{Smith2015} is one able to actually obtain the level spacings of a generic quantum many-body system. In that sense, alternative methods have been used to probe these phases. The most common is to follow the dynamic properties of a carefully prepared quantum system. Specifically, experiments in optical lattices employ high-fidelity preparation of initial states whose properties are well known, for example, by confining the atoms in certain regions of the trapping environment. By measuring how the information of the initial prepared state is preserved for long times after the release of this constraint one is able to identify the regimes where disorder is sufficient to lead to the MBL phenomenon. Thus, like for the experiments, we compute the unitary time evolution, governed by $\hat H = \hat H_0 + \hat H_{I}$, of a particular initial state: $|\Psi_0\rangle = |1 0 1 0 1 0\ldots\rangle$, which represents a fully formed charge-density wave. An observable that quantifies this initial information is called the charge imbalance ${\cal I}=(\langle\hat n^{\rm e}\rangle-\langle\hat n^{\rm o}\rangle)/(\langle\hat n^{\rm e}\rangle+\langle\hat n^{\rm o}\rangle)$, where $\hat n^{\rm e(o)}=\sum_{i=\text{even}(\text{odd})}\hat n_{i}$.

\begin{figure}[!tb] %%% FIG3
 \includegraphics[width=0.99\columnwidth]{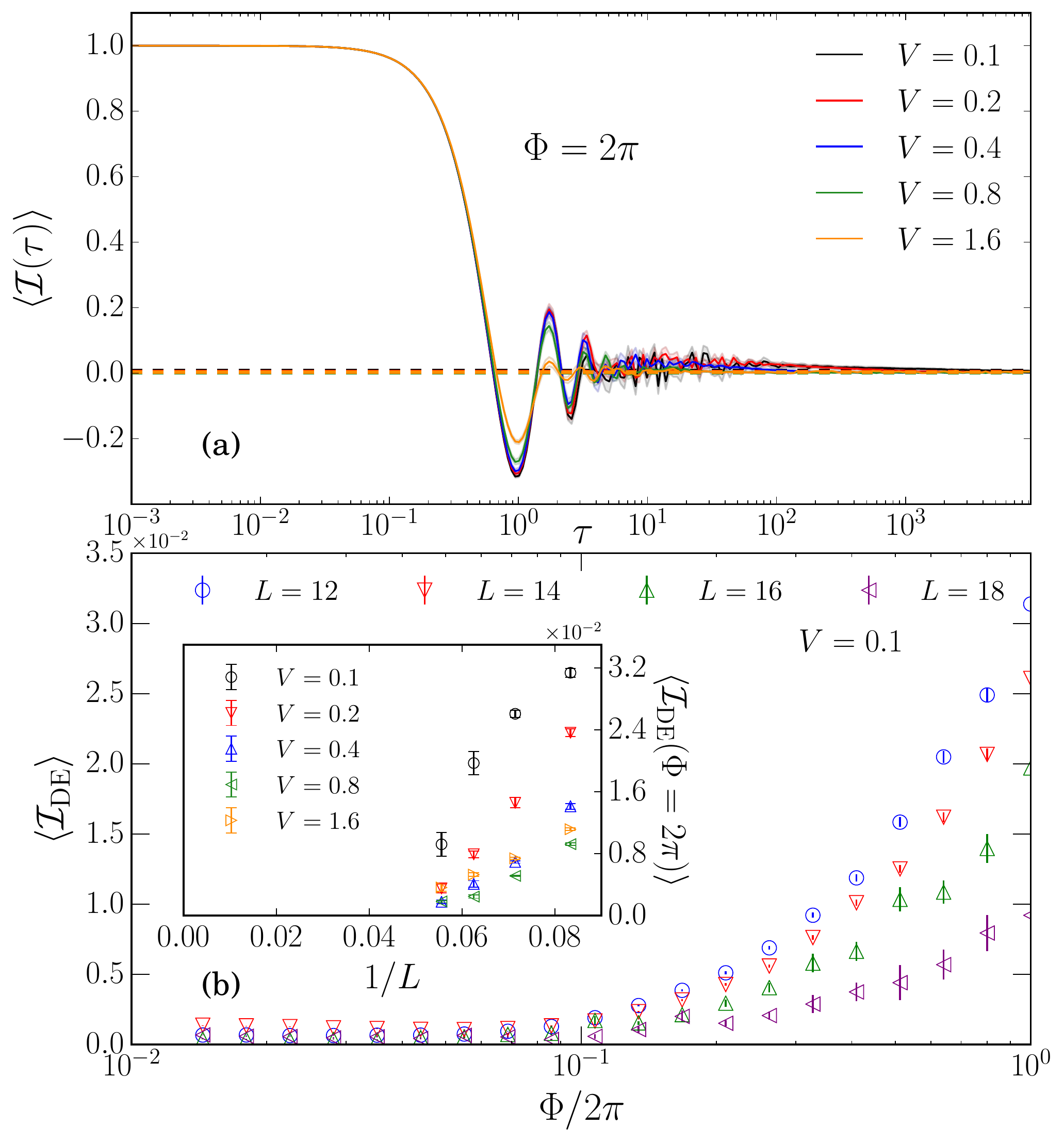}
 \vspace{-0.5cm}
 \caption{(Color online) (a) Relaxation dynamics of the imbalance in the disordered environment for $L=18$ and different interaction strengths $V$ with the largest possible disorder $\Phi=2\pi$. The amplitude of the coherent fluctuations around the equilibrium value (dashed lines) for small times is inversely proportional to the value of $V$.  (b) Diagonal ensemble results of the same quantity for $V = 0.1$; the inset shows the finite-size scaling of the DE prediction at the largest disorder strength, showing that in the thermodynamic limit the initial charge-density-wave information is lost in all cases. All the results are disorder averaged, and error bars [shaded areas in panel (a)] represent the standard deviation of the mean.}
 \label{fig:imbalance_time_DE}
\end{figure}

Figure~\ref{fig:imbalance_time_DE}(a) shows the relaxation dynamics of the imbalance for the largest disorder strength ($\Phi=2\pi$) and $L=18$; dashed lines depict the so-called diagonal ensemble (DE) values which for a system that does not possess degeneracies represent the infinite-time average for this observable~\cite{Rigol2008}. This latter quantity can be thought of as an order parameter of the MBL transition itself~\cite{Schreiber2015,Choi2016}. If the system thermalizes, the initial charge-density wave is washed out~\cite{Schreiber2015,Mondaini2015}, resulting in a vanishing imbalance in the infinite-time limit, whereas in the MBL phase it is markedly finite~\cite{Serbym2014b}. The DE prediction of the imbalance can be written in terms of the eigenstates $|\alpha\rangle$ with corresponding eigenenergies $E_\alpha$ of $\hat H$, the projections of the initial state in the eigenstates $c_\alpha = \langle\Psi_0|\alpha\rangle$, and the eigenstate expectation values ${\cal I}_{\alpha\alpha}$ via ${\cal I}_{\rm DE} \equiv \sum_\alpha |c_\alpha|^2{\cal I}_{\alpha\alpha}$~\cite{Rigol2008}. Consistent with the analysis of the ratio of adjacent gaps, the initial information ($\langle\Psi_0| {\cal \hat I}|\Psi_0\rangle = 1$) is lost after the time evolution with typical equilibration times \footnote{The Hamiltonian in the present form still possess further symmetries which are removed by inserting small symmetry breaking fields (see Appendix B).} shorter for larger strengths of the interactions. This suggests that the MBL phase is not present even when the disorder strength is maximum. A careful finite-size analysis is in order to support this claim. Figure ~\ref{fig:imbalance_time_DE}(b) displays the disorder average of ${\cal I}_{\rm DE}$ as a function of the disorder strength for different system sizes with the smallest interaction studied ($V=0.1$). In this limit, finite-size effects are larger for $\Phi=2\pi$, but system size extrapolation of these results (inset) shows that \textit{all} the initial information is lost at arbitrarily long times in the thermodynamic limit.

\begin{figure}[h]
  \centering
  \includegraphics[width=0.8\columnwidth]{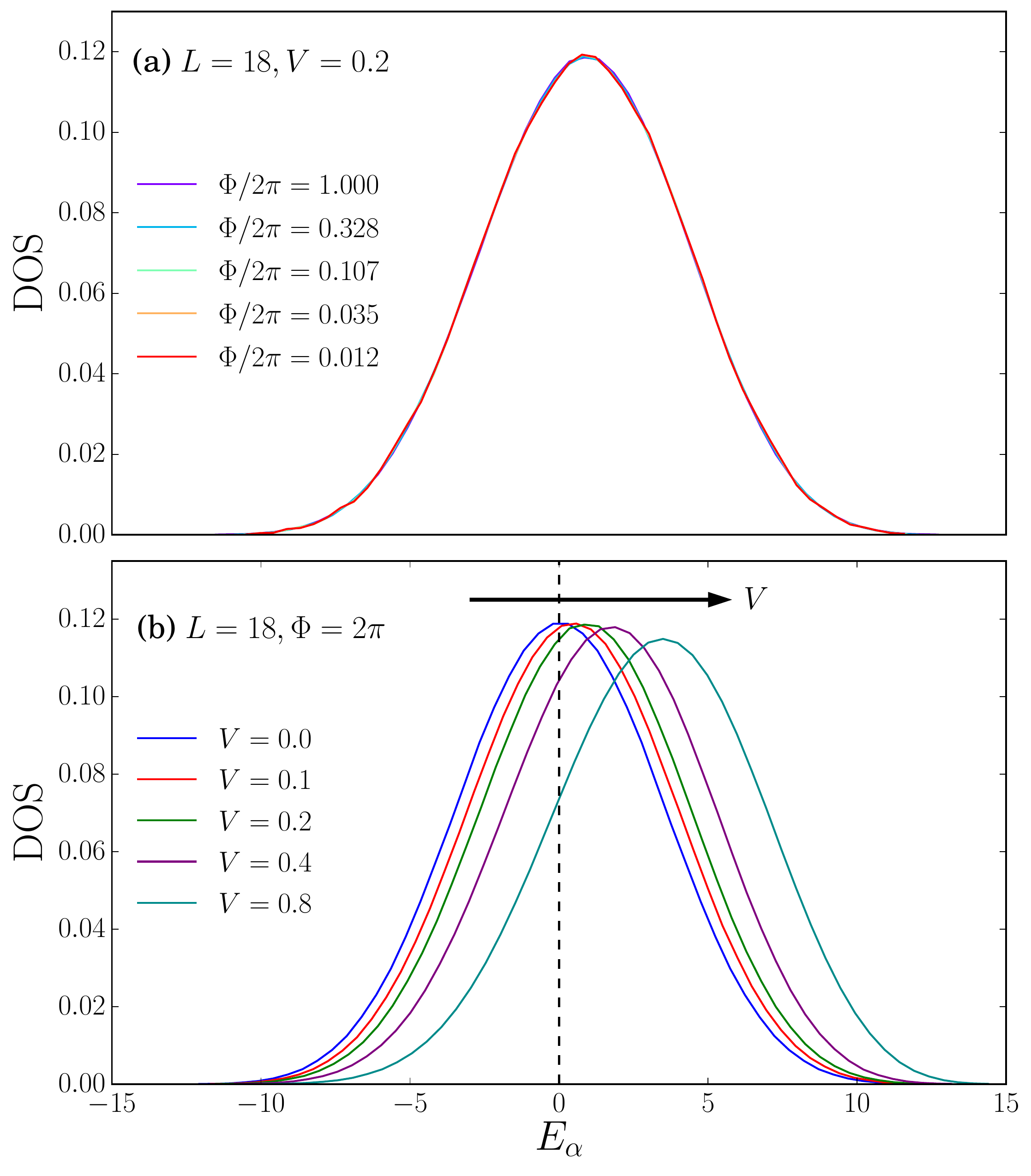}
  \caption{ The density of states in the many-body basis for $L=18$ (a) for fixed $V=0.2$ and increasing disorder $\Phi$ and (b) for $\Phi=2\pi$ and different interactions. In the latter case, larger interactions shift the center of the spectrum to higher energies. The vertical dashed line depicts the energy of the initial charge-density-wave state used in the quench dynamics.}
  \label{fig:int_DOS}
\end{figure}

The relaxation dynamics protocol mentioned above can be ill conditioned in experiments if the initial state is close to an eigenstate of the system. For this purpose, we investigate the many-body DOS of the interacting system, which is known to display a Gaussian shape. As an example, we plot the DOS for $L=18$, $V=0.2$, and several disorder strengths in Fig.~\ref{fig:int_DOS}(a). The inclusion of $\Phi$ does not substantially affect the DOS except at the edges of the spectrum. In turn, in Fig.~\ref{fig:int_DOS}(b) we show the dependence of the DOS with increasing interactions $V$. In this case the DOS shifts to higher energies as $V$ increases, but the energy of the initial state $\langle \Psi_0|\hat H|\Psi_0\rangle=0$ regardless of the interaction magnitude. Hence, in the large-$V$ limit, the initial state is dangerously close in energy to the ground state of the final Hamiltonian, and the proposed analysis one could use in experiments to probe the delocalization, i.e., by means of the quench dynamics of an initial charge-density-wave state, might be compromised in this limit. Nevertheless, the choice of $|\Psi_0\rangle$ in the present work is still able to demonstrate the delocalization for weak interactions.

\section{Thermalization and ETH predictions}

\begin{figure}[!tb] %%% FIG4
 \includegraphics[width=0.95\columnwidth]{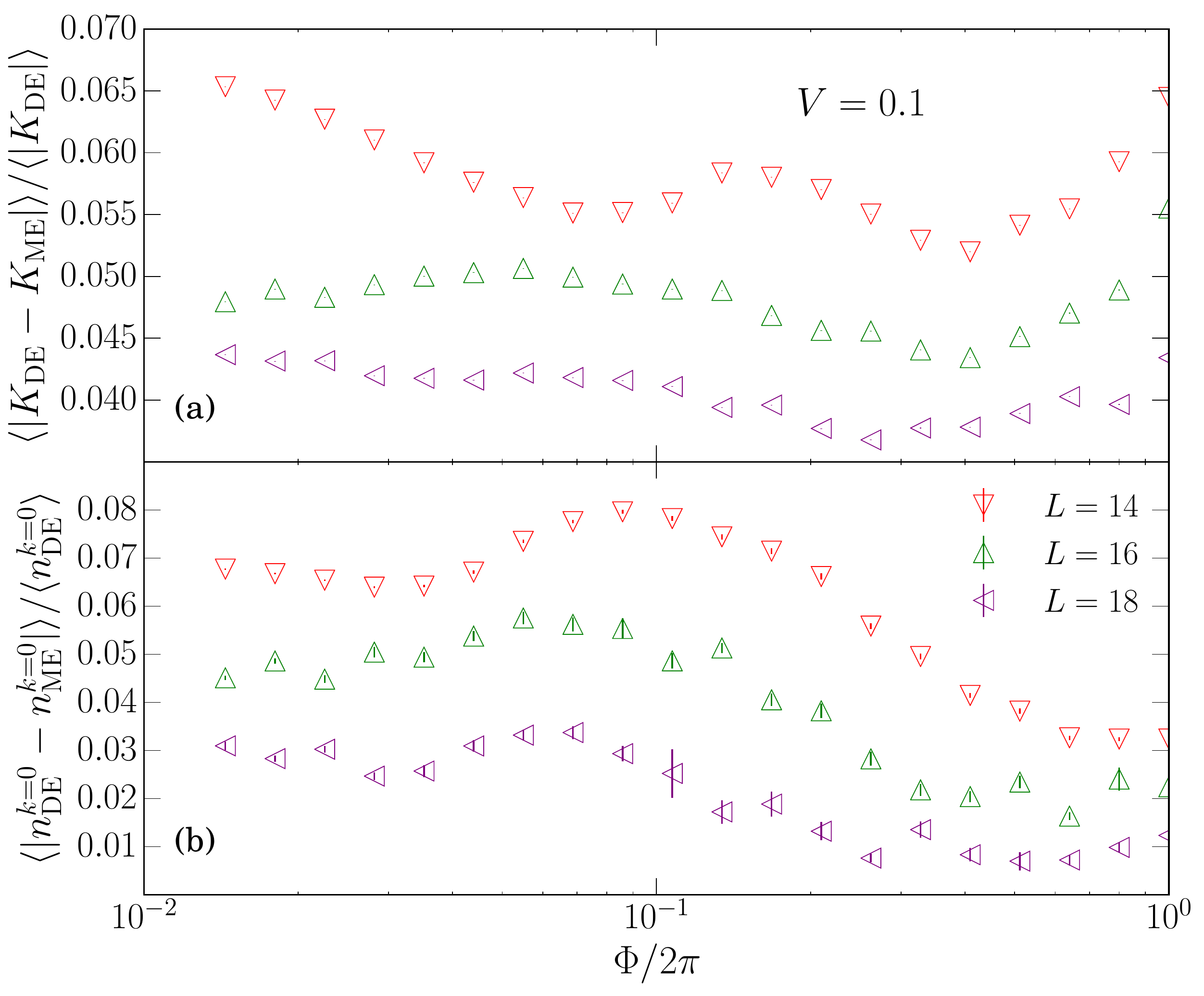}
 \vspace{-0.5cm}
 \caption{(Color online) Normalized disorder average difference between the diagonal and the microcanonical ensemble predictions vs the disorder strength $\Phi$ for (a) the kinetic energy and (b) the zero-momentum occupancy. In both cases we report the results for the smallest interactions studied, $V=0.1$, and the energy window for obtaining the microcanonical results is $\Delta E = 0.1$.}
 \label{fig:Fig_4_V01}
\end{figure}

The complete lack of information on the initial conditions after the relaxation dynamics is a characteristic of generic quantum systems that thermalize~\cite{Rigol2008,Dalessio2016}. In this situation the ETH predicts that the results of the infinite-time average and the estimates from the relevant statistical ensemble (for our isolated quantum system, we use the microcanonical one~\cite{rigol_09a}) should be equivalent in the thermodynamic limit if the system equilibrates [as we have verified in Fig.~\ref{fig:imbalance_time_DE}(a)]. To quantify this equivalence we show in Fig.~\ref{fig:Fig_4_V01} the relative difference between the predictions of the diagonal and microcanonical ensembles for two few-body observables: the kinetic energy ($\hat K = \hat H_0$) and the zero-momentum occupancy [$\hat n^{k=0} = (1/L)\sum_{i,j}\hat c^\dagger_i \hat c^{}_j$] for $V=0.1$ . The latter is readily obtained from absorption images after time-of-flight expansions in optical lattices~\cite{Bloch2008}. In both cases this relative difference is always $\lesssim8\%$ and decreases with the size of the system for the whole range of disorder values studied. A precise finite-size scaling with the few system sizes available is elusive, but this trend suggests that the prediction of the two ensembles will match in the thermodynamic limit. This ultimately leads to thermalization, as pointed out by the ETH, even for values of interactions $V$ of the order of 2\% of the noninteracting bandwidth.

\begin{figure}[!tb] %%% FIG5
 \includegraphics[width=0.99\columnwidth]{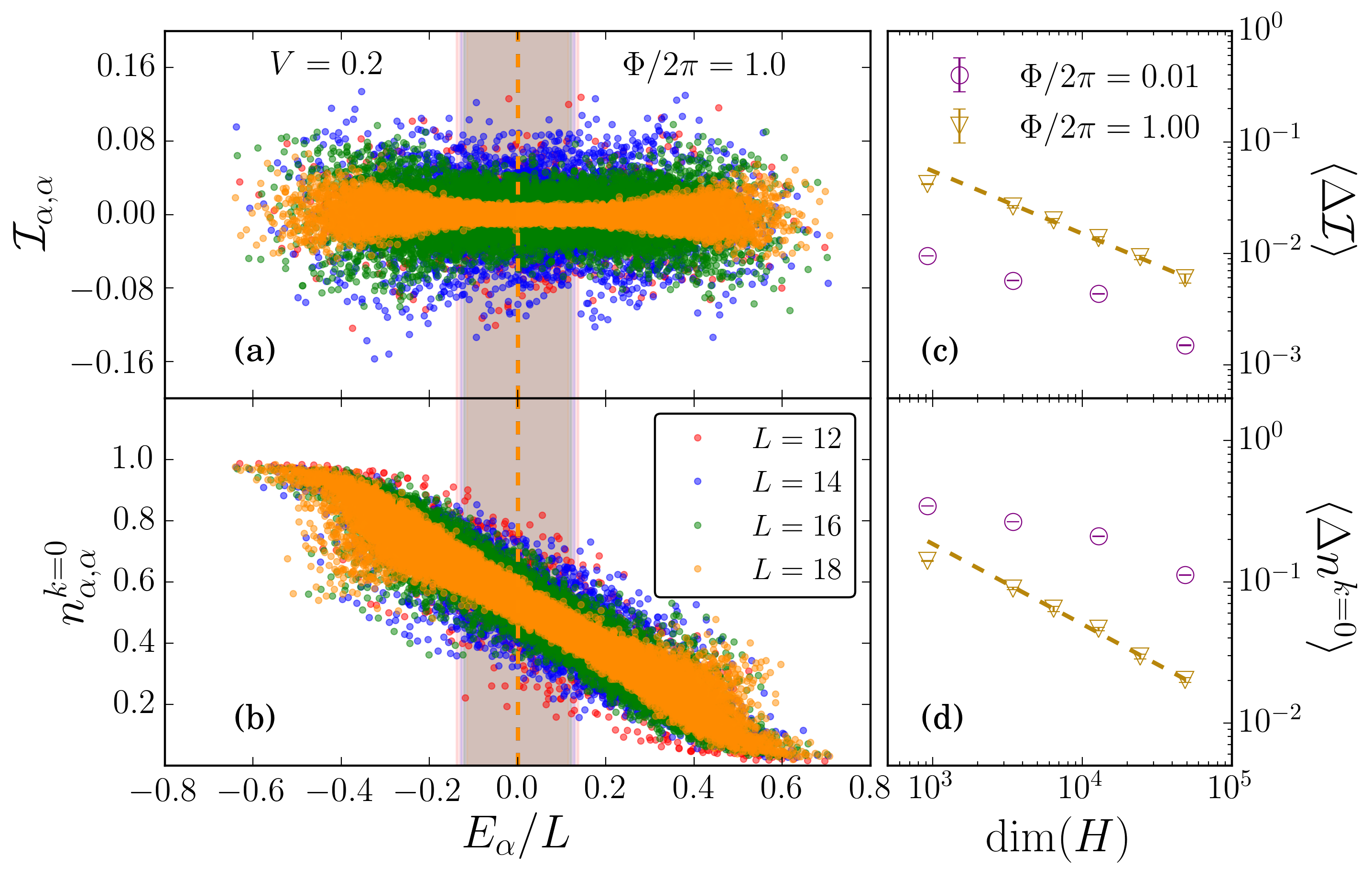}
 \vspace{-0.5cm}
 \caption{(Color online) Eigenstate expectation values of (a) the imbalance and (b) zero-momentum occupancy for the largest disorder amplitude $\Phi = 2\pi$ and $V=0.2$ for a single disorder realization vs eigenstate energy. The vertical dashed lines correspond to the total energy of the system, $\langle \hat H\rangle = \langle \Psi_0|\hat H|\Psi_0\rangle = 0$, and the shaded areas depict the corresponding values of initial-state variance, $\sigma = (\langle \hat H^2\rangle - \langle \hat H\rangle^2)^{1/2}$. (c) and (d) The fluctuations of consecutive eigenstate expectation values in an energy window defined by $x_{\rm thr.}=0.8$ as a function of the Hilbert space dimension for the corresponding observables displayed in (a) and (b). Dashed lines show the fitting to ${\rm dim}(H)^{a}$ for $L=14,15,16,17,18$. The lattice sizes with an odd number of sites have an occupancy of one extra particle on top of the half-filling density.}
 \label{fig:Fig_5_V02}
\end{figure}

Finally, ETH also predicts that the eigenstate expectation values of few-body operators in quantum systems that thermalize become a smooth function of the energy~\cite{Deutsch1991,Srednicki1994,Rigol2008,Dalessio2016}. We verify this scenario for the imbalance [Fig.~\ref{fig:Fig_5_V02}(a)] and for the zero-momentum occupation [Fig.~\ref{fig:Fig_5_V02}(b)] for the smallest value of interaction that already displays ergodicity in our finite systems ($V=0.2$) and the largest disorder $\Phi=2\pi$. Both quantities follow the ETH prediction with supports that become narrower with increasing system size. Quantitatively, this is related to the exponentially small eigenstate-to-eigenstate fluctuations for the expectation values of few-body operators with increasing system sizes. For a generic operator $\hat O$, we define this eigenstate-to-eigenstate fluctuation as $\Delta O \equiv |O_{\alpha\alpha} - O_{\alpha+1,\alpha+1}|$~\cite{beugeling_moessner_14,Kim2014,Mondaini2016}. In order to properly compute the average value of the fluctuations one needs to filter out the states at the edges of the spectrum, which do not exhibit quantum chaotic behavior in finite lattices with two-body interactions. On the other hand, it is not justifiable to select the states in the central half of the spectrum, as the corresponding range in energy becomes increasingly small for larger lattices due to the strong divergence of the DOS in the thermodynamic limit. Hence, to probe whether ETH works at temperatures other than \textit{infinite temperatures}, we select the eigenstates with eigenenergies $E_\alpha$ such that $(E_\alpha-E_{\rm GS})/|E_{\rm GS}|<x_\text{thr}$ and $(E_{\mathcal D}-E_{\alpha})/E_{\mathcal D}<x_\text{thr}$, where $E_{\rm GS}$ is the ground-state energy and $E_{\mathcal D}$ is the energy of the state of highest energy in the spectrum. In the present work we use $x_\text{thr}= 0.8$.

Figures~\ref{fig:Fig_5_V02}(c) and \ref{fig:Fig_5_V02}(d) display the dependence on the dimension of the Hilbert space $\text{dim}(H)$ of the disorder-averaged values for $\Delta {\cal I}$ and $\Delta n^{k=0}$, with two limit values of the disorder strength: $\Phi/2\pi = 0.01$ and $1$ for $V=0.2$. For $\Phi = 2\pi$ the fluctuations are seen to decay as a power law in $\text{dim}(H)$, suggesting that thermalization takes place even when the system is maximally disordered. The fitting power-law exponent $a=-0.53\pm0.02$ ($-0.55\pm0.02$) for $\langle\Delta{\cal I}\rangle$ ($\langle\Delta n^{k=0}\rangle$) is consistent with the results of Ref.~\cite{beugeling_moessner_14}, which shows that the exponent $a\approx -1/2$ is a characteristic of thermalizing systems.

\section{Summary and discussion.}
We investigated the ergodic properties of a quasi-one-dimensional chain subjected to disorder via a random vector potential in both the noninteracting and interacting cases. While in the former any arbitrarily small disorder induces single-particle localization, in the latter, small values of interactions are sufficient to lead to delocalization even for the largest possible disorder strength. We cannot claim that negligibly small interactions induce ergodicity since this is the limit in which finite-size effects are the largest. Nevertheless, the perturbative calculations of Ref.~\cite{Nandkishore2014} indicate that this scenario is expected for Hamiltonians which display marginal Anderson localization in the noninteracting limit whose localization length critical exponent $\nu$ is larger than 1 in one-dimensional systems. That is precisely the case here [see Fig.~\ref{fig:Fig_1_tprime05}(c)], thus suggesting that any finite interactions will lead to delocalization, regardless of the disorder strength in systems with quenched disorder modeled as a random vector potential. We conclude by stressing that this delocalization scenario may be investigated with current state-of-the-art techniques in optical lattices experiments.

\section*{Acknowledgments}
We acknowledge early discussions of this work with A. Saraiva and comments and suggestions from E. V. Castro, P. Sacramento, R. Nandkishore, D. J. Luitz, and M. Rigol. The computations were performed at the Institute for CyberScience at Penn State, the Center for High-Performance Computing at the University of Southern California, the Center of Interdisciplinary Studies in Lanzhou University, and the Tianhe-2JK at the Beijing Computational Science Research Center (CSRC). This research is financially supported by the National Natural Science Foundation of China (NSFC; Grants No. U1530401 and 11674021). R.M. also acknowledges support from NSFC (Grant No. 11650110441).

\appendix
\section{Determining the critical exponent $\nu$ for the noninteracting case}
\begin{figure}[ht]
  \centering
  \includegraphics[width=0.9\columnwidth]{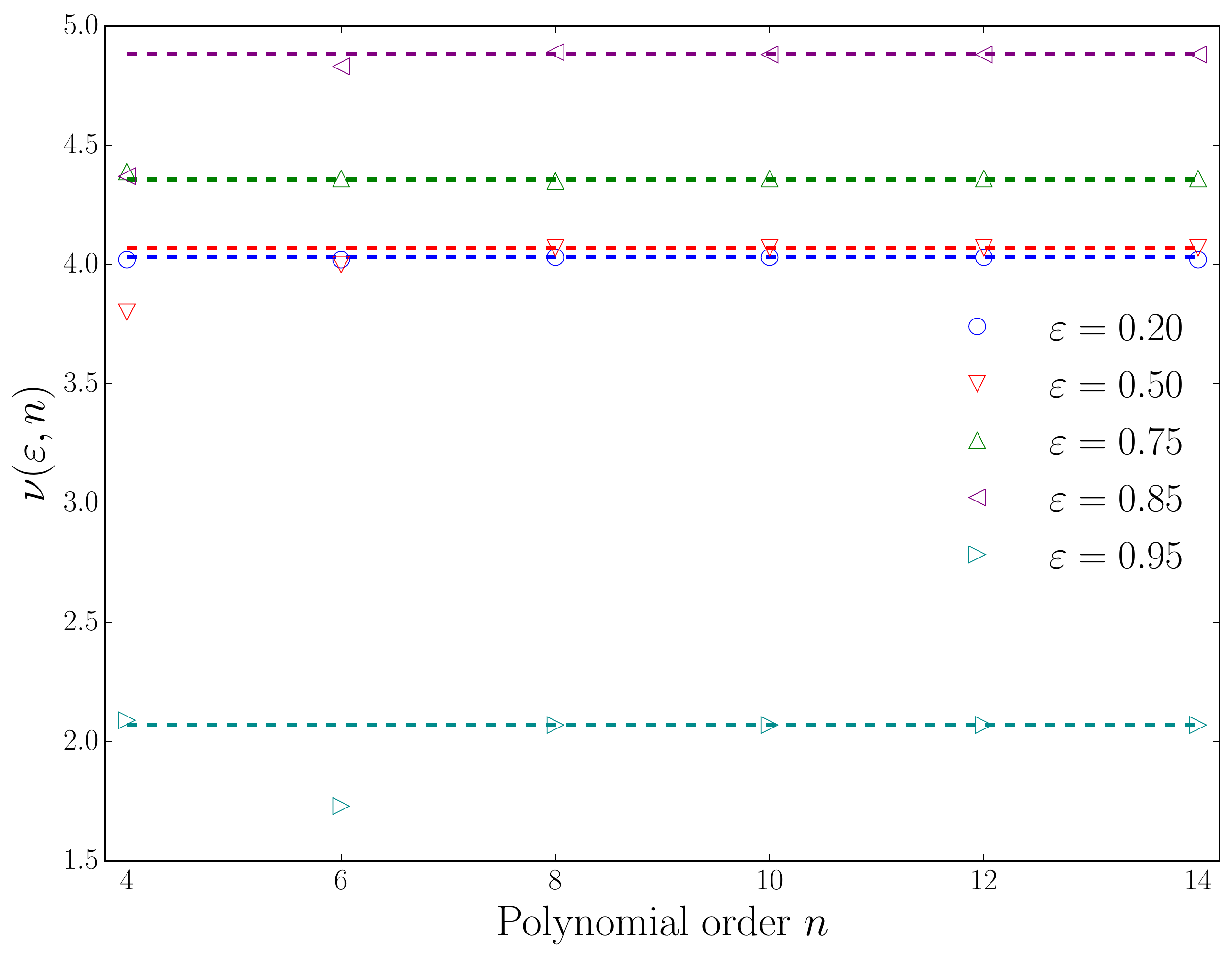}
  \caption{ (Color online) Best values of $\nu(\varepsilon,n)$ from least-squares fitting using different polynomial orders for different energy densities. Dashed lines denote $\nu(\varepsilon)=\frac{1}{4}\sum_{n=8,10,12,14}\nu(\varepsilon,n)$.
  }
  \label{fig:best_nu}
\end{figure}

Since the best value $\Phi_c(\varepsilon,n)$ is always zero for the different polynomial orders $n$ and different energy densities investigated, we can easily conclude $\Phi_c=0$ in the thermodynamic limit; that is, any finite disorder in the random vector potentials promotes localization in the single-particle picture. On the other hand, $\nu(\varepsilon,n)$ is more sensitive to the order of the polynomial used for the fittings. As an example, in Fig.~\ref{fig:best_nu} we systematically investigate how the exponent $\nu$ that scales the data depends on $n$ for different energy densities. The results for larger polynomial orders display a remarkable convergence; therefore, we determine the energy-density-dependent critical exponent $\nu$ [Fig.~1(c)] by the average of $\nu(\varepsilon,n)$ for the four largest polynomial orders.

\section{Symmetry-breaking field}
The $t-t^\prime-V$ model is nonintegrable for any nonzero $t'$ and displays Wigner-Dyson statistics for the level spacings in the thermodynamic limit. However, it is hard to numerically observe ergodicity (and associated thermalization) due to finite-size effects. In finite systems this is potentially compromised by extra symmetries which hinder the associated level repulsion. In the systems we study, with open boundary conditions, a mirror real-space symmetry is still present. To make sure the investigation of thermalization-localization transition starts from a thermal phase in the absence of disorder, we remove this symmetry by adding a local chemical potential term $\hat H_{\rm sb} = \mu_0 \hat n_0$ at one of the boundaries of the system.

\begin{figure}[ht]
  \centering
  \includegraphics[width=0.99\columnwidth]{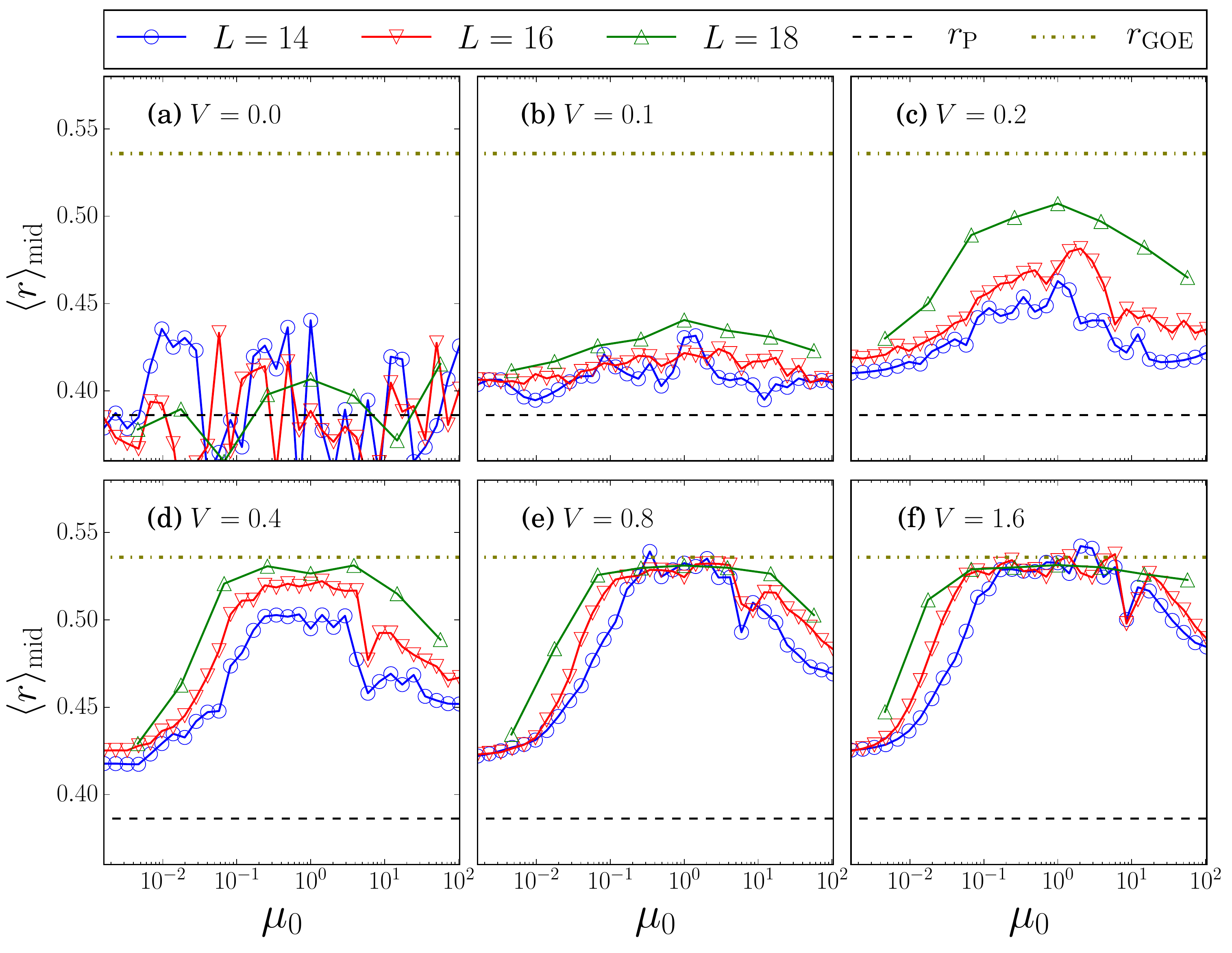}
  \caption{ Average ratio of adjacent gaps in the middle half of the spectrum as a function of $\mu_0$ for different interactions $V$ and system sizes $L$ in the absence of disorder.    }
  \label{fig:mu0test}
\end{figure}

We plot the mean value of the ratio of adjacent gaps in the middle half of the spectrum in Fig.~\ref{fig:mu0test} as a function of the local chemical potential strength $\mu_0$ for different interactions $V$ and system sizes $L$. While this chemical potential term is not relevant in the thermodynamic limit (since it is not extensive) the finite-size effects are largely reduced, especially for small system sizes, helping the average ratio of adjacent gaps to reach values closer to $r_{\rm GUE}$ in the clean system. In all the results presented for the interacting case in this work, we choose $\mu_0=0.1$.

\section{Distribution of the ratio of adjacent gaps}
\begin{figure}[!tb]
  \centering
  \includegraphics[width=0.99\columnwidth]{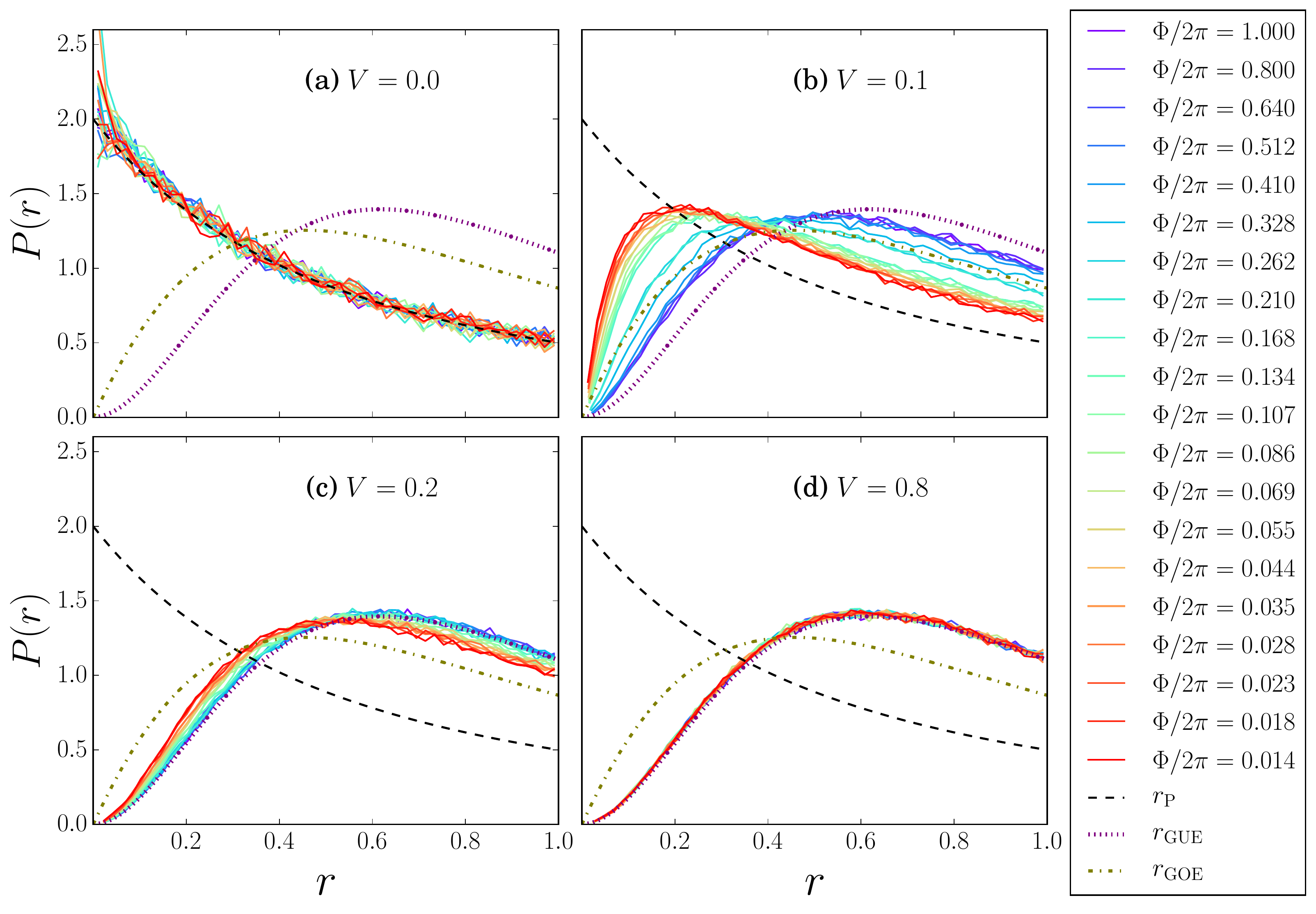}
  \caption{ Distribution of the ratio of adjacent gaps for $L=18$ and increasing interactions $V$ and different disorder strengths $\Phi$. For small interactions [see, e.g., (b)], the distributions for larger disorder values are systematically closer to the GUE ensemble prediction.}
  \label{fig:P_r}
\end{figure}

Finally, we show in Fig.~\ref{fig:P_r} the ratio of adjacent gaps for increasing interactions $V$ for the largest lattice where we use full diagonalization ($L=18$). One can see that the Poisson distribution is obtained only in the non-interacting limit (associated with localization) and, for finite interactions, the distributions depart from this limit to essentially become equivalent to the GUE distribution for moderately large values of $V$. The increasing values of disorder help in reducing the finite-size effects for small interaction values, and the largest disorder amplitudes get closer to the distribution for ergodic systems in comparison to smaller ones.

\bibliography{mbl_1d_random_vec_pot}

%merlin.mbs apsrev4-1.bst 2010-07-25 4.21a (PWD, AO, DPC) hacked
%Control: key (0)
%Control: author (0) dotless jnrlst
%Control: editor formatted (1) identically to author
%Control: production of article title (0) allowed
%Control: page (1) range
%Control: year (0) verbatim
%Control: production of eprint (0) enabled
\begin{thebibliography}{57}%
\makeatletter
\providecommand \@ifxundefined [1]{%
 \@ifx{#1\undefined}
}%
\providecommand \@ifnum [1]{%
 \ifnum #1\expandafter \@firstoftwo
 \else \expandafter \@secondoftwo
 \fi
}%
\providecommand \@ifx [1]{%
 \ifx #1\expandafter \@firstoftwo
 \else \expandafter \@secondoftwo
 \fi
}%
\providecommand \natexlab [1]{#1}%
\providecommand \enquote  [1]{``#1''}%
\providecommand \bibnamefont  [1]{#1}%
\providecommand \bibfnamefont [1]{#1}%
\providecommand \citenamefont [1]{#1}%
\providecommand \href@noop [0]{\@secondoftwo}%
\providecommand \href [0]{\begingroup \@sanitize@url \@href}%
\providecommand \@href[1]{\@@startlink{#1}\@@href}%
\providecommand \@@href[1]{\endgroup#1\@@endlink}%
\providecommand \@sanitize@url [0]{\catcode `\\12\catcode `\$12\catcode
  `\&12\catcode `\#12\catcode `\^12\catcode `\_12\catcode `\%12\relax}%
\providecommand \@@startlink[1]{}%
\providecommand \@@endlink[0]{}%
\providecommand \url  [0]{\begingroup\@sanitize@url \@url }%
\providecommand \@url [1]{\endgroup\@href {#1}{\urlprefix }}%
\providecommand \urlprefix  [0]{URL }%
\providecommand \Eprint [0]{\href }%
\providecommand \doibase [0]{http://dx.doi.org/}%
\providecommand \selectlanguage [0]{\@gobble}%
\providecommand \bibinfo  [0]{\@secondoftwo}%
\providecommand \bibfield  [0]{\@secondoftwo}%
\providecommand \translation [1]{[#1]}%
\providecommand \BibitemOpen [0]{}%
\providecommand \bibitemStop [0]{}%
\providecommand \bibitemNoStop [0]{.\EOS\space}%
\providecommand \EOS [0]{\spacefactor3000\relax}%
\providecommand \BibitemShut  [1]{\csname bibitem#1\endcsname}%
\let\auto@bib@innerbib\@empty
%</preamble>
\bibitem [{\citenamefont {Anderson}(1958)}]{Anderson1958}%
  \BibitemOpen
  \bibfield  {author} {\bibinfo {author} {\bibfnamefont {P.~W.}\ \bibnamefont
  {Anderson}},\ }\bibfield  {title} {\enquote {\bibinfo {title} {Absence of
  diffusion in certain random lattices},}\ }\href@noop {} {\bibfield  {journal}
  {\bibinfo  {journal} {Phys. Rev.}\ }\textbf {\bibinfo {volume} {109}},\
  \bibinfo {pages} {1492} (\bibinfo {year} {1958})}\BibitemShut {NoStop}%
\bibitem [{\citenamefont {Evers}\ and\ \citenamefont
  {Mirlin}(2008)}]{Evers_Mirlin_2008}%
  \BibitemOpen
  \bibfield  {author} {\bibinfo {author} {\bibfnamefont {F.}~\bibnamefont
  {Evers}}\ and\ \bibinfo {author} {\bibfnamefont {A.~D.}\ \bibnamefont
  {Mirlin}},\ }\bibfield  {title} {\enquote {\bibinfo {title} {Anderson
  transitions},}\ }\href@noop {} {\bibfield  {journal} {\bibinfo  {journal}
  {Rev. of Mod. Phys.}\ }\textbf {\bibinfo {volume} {80}},\ \bibinfo {pages}
  {1355} (\bibinfo {year} {2008})}\BibitemShut {NoStop}%
\bibitem [{\citenamefont {Abrahams}\ \emph {et~al.}(1979)\citenamefont
  {Abrahams}, \citenamefont {Anderson}, \citenamefont {Licciardello},\ and\
  \citenamefont {Ramakrishnan}}]{Abrahams1979}%
  \BibitemOpen
  \bibfield  {author} {\bibinfo {author} {\bibfnamefont {E.}~\bibnamefont
  {Abrahams}}, \bibinfo {author} {\bibfnamefont {P.~W.}\ \bibnamefont
  {Anderson}}, \bibinfo {author} {\bibfnamefont {D.~C.}\ \bibnamefont
  {Licciardello}}, \ and\ \bibinfo {author} {\bibfnamefont {T.~V.}\
  \bibnamefont {Ramakrishnan}},\ }\bibfield  {title} {\enquote {\bibinfo
  {title} {Scaling theory of localization: {A}bsence of quantum diffusion in
  two dimensions},}\ }\href@noop {} {\bibfield  {journal} {\bibinfo  {journal}
  {Phys. Rev. Lett.}\ }\textbf {\bibinfo {volume} {42}},\ \bibinfo {pages}
  {673--676} (\bibinfo {year} {1979})}\BibitemShut {NoStop}%
\bibitem [{\citenamefont {Fleishman}\ and\ \citenamefont
  {Anderson}(1980)}]{Fleishman80}%
  \BibitemOpen
  \bibfield  {author} {\bibinfo {author} {\bibfnamefont {L.}~\bibnamefont
  {Fleishman}}\ and\ \bibinfo {author} {\bibfnamefont {P.~W.}\ \bibnamefont
  {Anderson}},\ }\bibfield  {title} {\enquote {\bibinfo {title} {Interactions
  and the {A}nderson transition},}\ }\href {\doibase 10.1103/PhysRevB.21.2366}
  {\bibfield  {journal} {\bibinfo  {journal} {Phys. Rev. B}\ }\textbf {\bibinfo
  {volume} {21}},\ \bibinfo {pages} {2366--2377} (\bibinfo {year}
  {1980})}\BibitemShut {NoStop}%
\bibitem [{\citenamefont {Altshuler}\ \emph {et~al.}(1997)\citenamefont
  {Altshuler}, \citenamefont {Gefen}, \citenamefont {Kamenev},\ and\
  \citenamefont {Levitov}}]{Altshuler_Gefen_97}%
  \BibitemOpen
  \bibfield  {author} {\bibinfo {author} {\bibfnamefont {B.~L.}\ \bibnamefont
  {Altshuler}}, \bibinfo {author} {\bibfnamefont {Y.}~\bibnamefont {Gefen}},
  \bibinfo {author} {\bibfnamefont {A.}~\bibnamefont {Kamenev}}, \ and\
  \bibinfo {author} {\bibfnamefont {L.~S.}\ \bibnamefont {Levitov}},\
  }\bibfield  {title} {\enquote {\bibinfo {title} {Quasiparticle lifetime in a
  finite system: {A} nonperturbative approach},}\ }\href {\doibase
  10.1103/PhysRevLett.78.2803} {\bibfield  {journal} {\bibinfo  {journal}
  {Phys. Rev. Lett.}\ }\textbf {\bibinfo {volume} {78}},\ \bibinfo {pages}
  {2803--2806} (\bibinfo {year} {1997})}\BibitemShut {NoStop}%
\bibitem [{\citenamefont {Gornyi}\ \emph {et~al.}(2005)\citenamefont {Gornyi},
  \citenamefont {Mirlin},\ and\ \citenamefont {Polyakov}}]{Gornyi_Mirlin_05}%
  \BibitemOpen
  \bibfield  {author} {\bibinfo {author} {\bibfnamefont {I.~V.}\ \bibnamefont
  {Gornyi}}, \bibinfo {author} {\bibfnamefont {A.~D.}\ \bibnamefont {Mirlin}},
  \ and\ \bibinfo {author} {\bibfnamefont {D.~G.}\ \bibnamefont {Polyakov}},\
  }\bibfield  {title} {\enquote {\bibinfo {title} {Interacting electrons in
  disordered wires: {A}nderson localization and low-${T}$ transport},}\ }\href
  {\doibase 10.1103/PhysRevLett.95.206603} {\bibfield  {journal} {\bibinfo
  {journal} {Phys. Rev. Lett.}\ }\textbf {\bibinfo {volume} {95}},\ \bibinfo
  {pages} {206603} (\bibinfo {year} {2005})}\BibitemShut {NoStop}%
\bibitem [{\citenamefont {Basko}\ \emph {et~al.}(2006)\citenamefont {Basko},
  \citenamefont {Aleiner},\ and\ \citenamefont {Altshuler}}]{Basko2006}%
  \BibitemOpen
  \bibfield  {author} {\bibinfo {author} {\bibfnamefont {D.~M.}\ \bibnamefont
  {Basko}}, \bibinfo {author} {\bibfnamefont {I.~L.}\ \bibnamefont {Aleiner}},
  \ and\ \bibinfo {author} {\bibfnamefont {B.~L.}\ \bibnamefont {Altshuler}},\
  }\bibfield  {title} {\enquote {\bibinfo {title} {Metal-insulator transition
  in a weakly interacting many-electron system with localized single-particle
  states},}\ }\href@noop {} {\bibfield  {journal} {\bibinfo  {journal} {Ann.
  Phys.}\ }\textbf {\bibinfo {volume} {321}},\ \bibinfo {pages} {1126--1205}
  (\bibinfo {year} {2006})}\BibitemShut {NoStop}%
\bibitem [{\citenamefont {Basko}\ \emph {et~al.}(2007)\citenamefont {Basko},
  \citenamefont {Aleiner},\ and\ \citenamefont {Altshuler}}]{Basko2008}%
  \BibitemOpen
  \bibfield  {author} {\bibinfo {author} {\bibfnamefont {D.~M.}\ \bibnamefont
  {Basko}}, \bibinfo {author} {\bibfnamefont {I.~L.}\ \bibnamefont {Aleiner}},
  \ and\ \bibinfo {author} {\bibfnamefont {B.~L.}\ \bibnamefont {Altshuler}},\
  }\bibfield  {title} {\enquote {\bibinfo {title} {On the many-body
  localization phenomena},}\ }in\ \href@noop {} {\emph {\bibinfo {booktitle}
  {Problems of Condensed Matter Physics: Quantum coherence phenomena in
  electron-hole and coupled matter-light systems}}},\ \bibinfo {editor} {edited
  by\ \bibinfo {editor} {\bibfnamefont {Alexei~L.}\ \bibnamefont {Ivanov}}\
  and\ \bibinfo {editor} {\bibfnamefont {Sergei~G.}\ \bibnamefont
  {Tikhodeev}}}\ (\bibinfo  {publisher} {Oxford University Press},\ \bibinfo
  {year} {2007})\BibitemShut {NoStop}%
\bibitem [{\citenamefont {Oganesyan}\ and\ \citenamefont
  {Huse}(2007)}]{Oganesyan_Huse_07}%
  \BibitemOpen
  \bibfield  {author} {\bibinfo {author} {\bibfnamefont {V.}~\bibnamefont
  {Oganesyan}}\ and\ \bibinfo {author} {\bibfnamefont {D.~A.}\ \bibnamefont
  {Huse}},\ }\bibfield  {title} {\enquote {\bibinfo {title} {Localization of
  interacting fermions at high temperature},}\ }\href {\doibase
  10.1103/PhysRevB.75.155111} {\bibfield  {journal} {\bibinfo  {journal} {Phys.
  Rev. B}\ }\textbf {\bibinfo {volume} {75}},\ \bibinfo {pages} {155111}
  (\bibinfo {year} {2007})}\BibitemShut {NoStop}%
\bibitem [{\citenamefont {\ifmmode \check{Z}\else
  \v{Z}\fi{}nidari\ifmmode~\check{c}\else \v{c}\fi{}}\ \emph
  {et~al.}(2008)\citenamefont {\ifmmode \check{Z}\else
  \v{Z}\fi{}nidari\ifmmode~\check{c}\else \v{c}\fi{}}, \citenamefont {Prosen},\
  and\ \citenamefont {Prelov\ifmmode~\check{s}\else
  \v{s}\fi{}ek}}]{Znidaric08}%
  \BibitemOpen
  \bibfield  {author} {\bibinfo {author} {\bibfnamefont {M.}~\bibnamefont
  {\ifmmode \check{Z}\else \v{Z}\fi{}nidari\ifmmode~\check{c}\else
  \v{c}\fi{}}}, \bibinfo {author} {\bibfnamefont {T.}~\bibnamefont {Prosen}}, \
  and\ \bibinfo {author} {\bibfnamefont {P.}~\bibnamefont
  {Prelov\ifmmode~\check{s}\else \v{s}\fi{}ek}},\ }\bibfield  {title} {\enquote
  {\bibinfo {title} {Many-body localization in the {H}eisenberg ${XXZ}$ magnet
  in a random field},}\ }\href {\doibase 10.1103/PhysRevB.77.064426} {\bibfield
   {journal} {\bibinfo  {journal} {Phys. Rev. B}\ }\textbf {\bibinfo {volume}
  {77}},\ \bibinfo {pages} {064426} (\bibinfo {year} {2008})}\BibitemShut
  {NoStop}%
\bibitem [{\citenamefont {Pal}\ and\ \citenamefont {Huse}(2010)}]{Pal10}%
  \BibitemOpen
  \bibfield  {author} {\bibinfo {author} {\bibfnamefont {A.}~\bibnamefont
  {Pal}}\ and\ \bibinfo {author} {\bibfnamefont {D.~A.}\ \bibnamefont {Huse}},\
  }\bibfield  {title} {\enquote {\bibinfo {title} {Many-body localization phase
  transition},}\ }\href {\doibase 10.1103/PhysRevB.82.174411} {\bibfield
  {journal} {\bibinfo  {journal} {Phys. Rev. B}\ }\textbf {\bibinfo {volume}
  {82}},\ \bibinfo {pages} {174411} (\bibinfo {year} {2010})}\BibitemShut
  {NoStop}%
\bibitem [{\citenamefont {Khatami}\ \emph {et~al.}(2012)\citenamefont
  {Khatami}, \citenamefont {Rigol}, \citenamefont {Rela\~no},\ and\
  \citenamefont {Garcia-Garcia}}]{Khatami_Rigol_12}%
  \BibitemOpen
  \bibfield  {author} {\bibinfo {author} {\bibfnamefont {E.}~\bibnamefont
  {Khatami}}, \bibinfo {author} {\bibfnamefont {M.}~\bibnamefont {Rigol}},
  \bibinfo {author} {\bibfnamefont {A.}~\bibnamefont {Rela\~no}}, \ and\
  \bibinfo {author} {\bibfnamefont {A.~M.}\ \bibnamefont {Garcia-Garcia}},\
  }\bibfield  {title} {\enquote {\bibinfo {title} {Quantum quenches in
  disordered systems: {A}pproach to thermal equilibrium without a typical
  relaxation time},}\ }\href {\doibase 10.1103/PhysRevE.85.050102} {\bibfield
  {journal} {\bibinfo  {journal} {Phys. Rev. E}\ }\textbf {\bibinfo {volume}
  {85}},\ \bibinfo {pages} {050102(R)} (\bibinfo {year} {2012})}\BibitemShut
  {NoStop}%
\bibitem [{\citenamefont {Bardarson}\ \emph {et~al.}(2012)\citenamefont
  {Bardarson}, \citenamefont {Pollmann},\ and\ \citenamefont
  {Moore}}]{Bardarson_Pollmann_12}%
  \BibitemOpen
  \bibfield  {author} {\bibinfo {author} {\bibfnamefont {J.~H.}\ \bibnamefont
  {Bardarson}}, \bibinfo {author} {\bibfnamefont {F.}~\bibnamefont {Pollmann}},
  \ and\ \bibinfo {author} {\bibfnamefont {J.~E.}\ \bibnamefont {Moore}},\
  }\bibfield  {title} {\enquote {\bibinfo {title} {Unbounded growth of
  entanglement in models of many-body localization},}\ }\href {\doibase
  10.1103/PhysRevLett.109.017202} {\bibfield  {journal} {\bibinfo  {journal}
  {Phys. Rev. Lett.}\ }\textbf {\bibinfo {volume} {109}},\ \bibinfo {pages}
  {017202} (\bibinfo {year} {2012})}\BibitemShut {NoStop}%
\bibitem [{\citenamefont {Luitz}\ \emph {et~al.}(2015)\citenamefont {Luitz},
  \citenamefont {Laflorencie},\ and\ \citenamefont {Alet}}]{Luitz2015}%
  \BibitemOpen
  \bibfield  {author} {\bibinfo {author} {\bibfnamefont {D.~J.}\ \bibnamefont
  {Luitz}}, \bibinfo {author} {\bibfnamefont {N.}~\bibnamefont {Laflorencie}},
  \ and\ \bibinfo {author} {\bibfnamefont {F.}~\bibnamefont {Alet}},\
  }\bibfield  {title} {\enquote {\bibinfo {title} {Many-body localization edge
  in the random-field {H}eisenberg chain},}\ }\href@noop {} {\bibfield
  {journal} {\bibinfo  {journal} {Phys. Rev. B}\ }\textbf {\bibinfo {volume}
  {91}},\ \bibinfo {pages} {081103} (\bibinfo {year} {2015})}\BibitemShut
  {NoStop}%
\bibitem [{\citenamefont {BarLev}\ \emph {et~al.}(2015)\citenamefont {BarLev},
  \citenamefont {Cohen},\ and\ \citenamefont {Reichman}}]{Lev2015a}%
  \BibitemOpen
  \bibfield  {author} {\bibinfo {author} {\bibfnamefont {Y.}~\bibnamefont
  {BarLev}}, \bibinfo {author} {\bibfnamefont {G.}~\bibnamefont {Cohen}}, \
  and\ \bibinfo {author} {\bibfnamefont {D.~R.}\ \bibnamefont {Reichman}},\
  }\bibfield  {title} {\enquote {\bibinfo {title} {Absence of diffusion in an
  interacting system of spinless fermions on a one-dimensional disordered
  lattice},}\ }\href@noop {} {\bibfield  {journal} {\bibinfo  {journal} {Phys.
  Rev. Lett.}\ }\textbf {\bibinfo {volume} {114}},\ \bibinfo {pages} {100601}
  (\bibinfo {year} {2015})}\BibitemShut {NoStop}%
\bibitem [{\citenamefont {Mondaini}\ and\ \citenamefont
  {Rigol}(2015)}]{Mondaini2015}%
  \BibitemOpen
  \bibfield  {author} {\bibinfo {author} {\bibfnamefont {R.}~\bibnamefont
  {Mondaini}}\ and\ \bibinfo {author} {\bibfnamefont {M.}~\bibnamefont
  {Rigol}},\ }\bibfield  {title} {\enquote {\bibinfo {title} {Many-body
  localization and thermalization in disordered {H}ubbard chains},}\
  }\href@noop {} {\bibfield  {journal} {\bibinfo  {journal} {Phys. Rev. A}\
  }\textbf {\bibinfo {volume} {92}},\ \bibinfo {pages} {041601} (\bibinfo
  {year} {2015})}\BibitemShut {NoStop}%
\bibitem [{\citenamefont {Nandkishore}\ and\ \citenamefont
  {Huse}(2015)}]{Nandkishore_Huse_review_15}%
  \BibitemOpen
  \bibfield  {author} {\bibinfo {author} {\bibfnamefont {R.}~\bibnamefont
  {Nandkishore}}\ and\ \bibinfo {author} {\bibfnamefont {D.~A.}\ \bibnamefont
  {Huse}},\ }\bibfield  {title} {\enquote {\bibinfo {title} {Many-body
  localization and thermalization in quantum statistical mechanics},}\
  }\href@noop {} {\bibfield  {journal} {\bibinfo  {journal} {Annual Review of
  Condensed Matter Physics}\ }\textbf {\bibinfo {volume} {6}},\ \bibinfo
  {pages} {15--38} (\bibinfo {year} {2015})}\BibitemShut {NoStop}%
\bibitem [{\citenamefont {Altman}\ and\ \citenamefont
  {Vosk}(2015)}]{Altman_Vosk_review_15}%
  \BibitemOpen
  \bibfield  {author} {\bibinfo {author} {\bibfnamefont {E.}~\bibnamefont
  {Altman}}\ and\ \bibinfo {author} {\bibfnamefont {R.}~\bibnamefont {Vosk}},\
  }\bibfield  {title} {\enquote {\bibinfo {title} {Universal dynamics and
  renormalization in many-body-localized systems},}\ }\href@noop {} {\bibfield
  {journal} {\bibinfo  {journal} {Annual Review of Condensed Matter Physics}\
  }\textbf {\bibinfo {volume} {6}},\ \bibinfo {pages} {383--409} (\bibinfo
  {year} {2015})}\BibitemShut {NoStop}%
\bibitem [{\citenamefont {Schreiber}\ \emph {et~al.}(2015)\citenamefont
  {Schreiber}, \citenamefont {Hodgman}, \citenamefont {Bordia}, \citenamefont
  {L{\"u}schen}, \citenamefont {Fischer}, \citenamefont {Vosk}, \citenamefont
  {Altman}, \citenamefont {Schneider},\ and\ \citenamefont
  {Bloch}}]{Schreiber2015}%
  \BibitemOpen
  \bibfield  {author} {\bibinfo {author} {\bibfnamefont {M.}~\bibnamefont
  {Schreiber}}, \bibinfo {author} {\bibfnamefont {S.~S.}\ \bibnamefont
  {Hodgman}}, \bibinfo {author} {\bibfnamefont {P.}~\bibnamefont {Bordia}},
  \bibinfo {author} {\bibfnamefont {Henrik~P.}\ \bibnamefont {L{\"u}schen}},
  \bibinfo {author} {\bibfnamefont {M.~H}\ \bibnamefont {Fischer}}, \bibinfo
  {author} {\bibfnamefont {R.}~\bibnamefont {Vosk}}, \bibinfo {author}
  {\bibfnamefont {E.}~\bibnamefont {Altman}}, \bibinfo {author} {\bibfnamefont
  {U.}~\bibnamefont {Schneider}}, \ and\ \bibinfo {author} {\bibfnamefont
  {I.}~\bibnamefont {Bloch}},\ }\bibfield  {title} {\enquote {\bibinfo {title}
  {Observation of many-body localization of interacting fermions in a
  quasirandom optical lattice},}\ }\href@noop {} {\bibfield  {journal}
  {\bibinfo  {journal} {Science}\ }\textbf {\bibinfo {volume} {349}},\ \bibinfo
  {pages} {842--845} (\bibinfo {year} {2015})}\BibitemShut {NoStop}%
\bibitem [{\citenamefont {Bordia}\ \emph {et~al.}(2016)\citenamefont {Bordia},
  \citenamefont {L{\"u}schen}, \citenamefont {Hodgman}, \citenamefont
  {Schreiber}, \citenamefont {Bloch},\ and\ \citenamefont
  {Schneider}}]{Bordia2016}%
  \BibitemOpen
  \bibfield  {author} {\bibinfo {author} {\bibfnamefont {P.}~\bibnamefont
  {Bordia}}, \bibinfo {author} {\bibfnamefont {H.~P.}\ \bibnamefont
  {L{\"u}schen}}, \bibinfo {author} {\bibfnamefont {S.~S.}\ \bibnamefont
  {Hodgman}}, \bibinfo {author} {\bibfnamefont {M.}~\bibnamefont {Schreiber}},
  \bibinfo {author} {\bibfnamefont {I.}~\bibnamefont {Bloch}}, \ and\ \bibinfo
  {author} {\bibfnamefont {U.}~\bibnamefont {Schneider}},\ }\bibfield  {title}
  {\enquote {\bibinfo {title} {Coupling identical one-dimensional many-body
  localized systems},}\ }\href@noop {} {\bibfield  {journal} {\bibinfo
  {journal} {Phys. Rev. Lett.}\ }\textbf {\bibinfo {volume} {116}},\ \bibinfo
  {pages} {140401} (\bibinfo {year} {2016})}\BibitemShut {NoStop}%
\bibitem [{\citenamefont {Choi}\ \emph {et~al.}(2016)\citenamefont {Choi},
  \citenamefont {Hild}, \citenamefont {Zeiher}, \citenamefont {Schau{\ss}},
  \citenamefont {Rubio-Abadal}, \citenamefont {Yefsah}, \citenamefont
  {Khemani}, \citenamefont {Huse}, \citenamefont {Bloch},\ and\ \citenamefont
  {Gross}}]{Choi2016}%
  \BibitemOpen
  \bibfield  {author} {\bibinfo {author} {\bibfnamefont {J.-Y.}\ \bibnamefont
  {Choi}}, \bibinfo {author} {\bibfnamefont {S.}~\bibnamefont {Hild}}, \bibinfo
  {author} {\bibfnamefont {J.}~\bibnamefont {Zeiher}}, \bibinfo {author}
  {\bibfnamefont {P.}~\bibnamefont {Schau{\ss}}}, \bibinfo {author}
  {\bibfnamefont {A.}~\bibnamefont {Rubio-Abadal}}, \bibinfo {author}
  {\bibfnamefont {T.}~\bibnamefont {Yefsah}}, \bibinfo {author} {\bibfnamefont
  {V.}~\bibnamefont {Khemani}}, \bibinfo {author} {\bibfnamefont {D.~A.}\
  \bibnamefont {Huse}}, \bibinfo {author} {\bibfnamefont {I.}~\bibnamefont
  {Bloch}}, \ and\ \bibinfo {author} {\bibfnamefont {C.}~\bibnamefont
  {Gross}},\ }\bibfield  {title} {\enquote {\bibinfo {title} {Exploring the
  many-body localization transition in two dimensions},}\ }\href@noop {}
  {\bibfield  {journal} {\bibinfo  {journal} {Science}\ }\textbf {\bibinfo
  {volume} {352}},\ \bibinfo {pages} {1547--1552} (\bibinfo {year}
  {2016})}\BibitemShut {NoStop}%
\bibitem [{\citenamefont {Smith}\ \emph {et~al.}(2016)\citenamefont {Smith},
  \citenamefont {Lee}, \citenamefont {Richerme}, \citenamefont {Neyenhuis},
  \citenamefont {Hess}, \citenamefont {Hauke}, \citenamefont {Heyl},
  \citenamefont {Huse},\ and\ \citenamefont {Monroe}}]{Smith2015}%
  \BibitemOpen
  \bibfield  {author} {\bibinfo {author} {\bibfnamefont {J.}~\bibnamefont
  {Smith}}, \bibinfo {author} {\bibfnamefont {A.}~\bibnamefont {Lee}}, \bibinfo
  {author} {\bibfnamefont {P.}~\bibnamefont {Richerme}}, \bibinfo {author}
  {\bibfnamefont {B.}~\bibnamefont {Neyenhuis}}, \bibinfo {author}
  {\bibfnamefont {P.~W.}\ \bibnamefont {Hess}}, \bibinfo {author}
  {\bibfnamefont {P.}~\bibnamefont {Hauke}}, \bibinfo {author} {\bibfnamefont
  {M.}~\bibnamefont {Heyl}}, \bibinfo {author} {\bibfnamefont {D.~A.}\
  \bibnamefont {Huse}}, \ and\ \bibinfo {author} {\bibfnamefont
  {C.}~\bibnamefont {Monroe}},\ }\bibfield  {title} {\enquote {\bibinfo {title}
  {Many-body localization in a quantum simulator with programmable random
  disorder},}\ }\href@noop {} {\bibfield  {journal} {\bibinfo  {journal} {Nat.
  Phys.}\ }\textbf {\bibinfo {volume} {12}},\ \bibinfo {pages} {907} (\bibinfo
  {year} {2016})}\BibitemShut {NoStop}%
\bibitem [{\citenamefont {Deutsch}(1991)}]{Deutsch1991}%
  \BibitemOpen
  \bibfield  {author} {\bibinfo {author} {\bibfnamefont {J.~M.}\ \bibnamefont
  {Deutsch}},\ }\bibfield  {title} {\enquote {\bibinfo {title} {Quantum
  statistical mechanics in a closed system},}\ }\href@noop {} {\bibfield
  {journal} {\bibinfo  {journal} {Phys. Rev. A}\ }\textbf {\bibinfo {volume}
  {43}},\ \bibinfo {pages} {2046} (\bibinfo {year} {1991})}\BibitemShut
  {NoStop}%
\bibitem [{\citenamefont {Srednicki}(1994)}]{Srednicki1994}%
  \BibitemOpen
  \bibfield  {author} {\bibinfo {author} {\bibfnamefont {M.}~\bibnamefont
  {Srednicki}},\ }\bibfield  {title} {\enquote {\bibinfo {title} {Chaos and
  quantum thermalization},}\ }\href@noop {} {\bibfield  {journal} {\bibinfo
  {journal} {Phys. Rev. E}\ }\textbf {\bibinfo {volume} {50}},\ \bibinfo
  {pages} {888} (\bibinfo {year} {1994})}\BibitemShut {NoStop}%
\bibitem [{\citenamefont {Rigol}\ \emph {et~al.}(2008)\citenamefont {Rigol},
  \citenamefont {Dunjko},\ and\ \citenamefont {Olshanii}}]{Rigol2008}%
  \BibitemOpen
  \bibfield  {author} {\bibinfo {author} {\bibfnamefont {M.}~\bibnamefont
  {Rigol}}, \bibinfo {author} {\bibfnamefont {V.}~\bibnamefont {Dunjko}}, \
  and\ \bibinfo {author} {\bibfnamefont {M.}~\bibnamefont {Olshanii}},\
  }\bibfield  {title} {\enquote {\bibinfo {title} {Thermalization and its
  mechanism for generic isolated quantum systems},}\ }\href@noop {} {\bibfield
  {journal} {\bibinfo  {journal} {Nature}\ }\textbf {\bibinfo {volume} {452}},\
  \bibinfo {pages} {854--858} (\bibinfo {year} {2008})}\BibitemShut {NoStop}%
\bibitem [{\citenamefont {D'Alessio}\ \emph {et~al.}(2016)\citenamefont
  {D'Alessio}, \citenamefont {Kafri}, \citenamefont {Polkovnikov},\ and\
  \citenamefont {Rigol}}]{Dalessio2016}%
  \BibitemOpen
  \bibfield  {author} {\bibinfo {author} {\bibfnamefont {L.}~\bibnamefont
  {D'Alessio}}, \bibinfo {author} {\bibfnamefont {Y.}~\bibnamefont {Kafri}},
  \bibinfo {author} {\bibfnamefont {A.}~\bibnamefont {Polkovnikov}}, \ and\
  \bibinfo {author} {\bibfnamefont {M.}~\bibnamefont {Rigol}},\ }\bibfield
  {title} {\enquote {\bibinfo {title} {From quantum chaos and eigenstate
  thermalization to statistical mechanics and thermodynamics},}\ }\href
  {http://dx.doi.org/10.1080/00018732.2016.1198134} {\bibfield  {journal}
  {\bibinfo  {journal} {Advances in Physics}\ }\textbf {\bibinfo {volume}
  {65}},\ \bibinfo {pages} {239--362} (\bibinfo {year} {2016})},\ \Eprint
  {http://arxiv.org/abs/http://dx.doi.org/10.1080/00018732.2016.1198134}
  {http://dx.doi.org/10.1080/00018732.2016.1198134} \BibitemShut {NoStop}%
\bibitem [{\citenamefont {Serbyn}\ \emph
  {et~al.}(2014{\natexlab{a}})\citenamefont {Serbyn}, \citenamefont {Knap},
  \citenamefont {Gopalakrishnan}, \citenamefont {Papi\ifmmode~\acute{c}\else
  \'{c}\fi{}}, \citenamefont {Yao}, \citenamefont {Laumann}, \citenamefont
  {Abanin}, \citenamefont {Lukin},\ and\ \citenamefont {Demler}}]{Serbym2014a}%
  \BibitemOpen
  \bibfield  {author} {\bibinfo {author} {\bibfnamefont {M.}~\bibnamefont
  {Serbyn}}, \bibinfo {author} {\bibfnamefont {M.}~\bibnamefont {Knap}},
  \bibinfo {author} {\bibfnamefont {S.}~\bibnamefont {Gopalakrishnan}},
  \bibinfo {author} {\bibfnamefont {Z.}~\bibnamefont
  {Papi\ifmmode~\acute{c}\else \'{c}\fi{}}}, \bibinfo {author} {\bibfnamefont
  {N.~Y.}\ \bibnamefont {Yao}}, \bibinfo {author} {\bibfnamefont {C.~R.}\
  \bibnamefont {Laumann}}, \bibinfo {author} {\bibfnamefont {D.~A.}\
  \bibnamefont {Abanin}}, \bibinfo {author} {\bibfnamefont {M.~D.}\
  \bibnamefont {Lukin}}, \ and\ \bibinfo {author} {\bibfnamefont {E.~A.}\
  \bibnamefont {Demler}},\ }\bibfield  {title} {\enquote {\bibinfo {title}
  {Interferometric probes of many-body localization},}\ }\href {\doibase
  10.1103/PhysRevLett.113.147204} {\bibfield  {journal} {\bibinfo  {journal}
  {Phys. Rev. Lett.}\ }\textbf {\bibinfo {volume} {113}},\ \bibinfo {pages}
  {147204} (\bibinfo {year} {2014}{\natexlab{a}})}\BibitemShut {NoStop}%
\bibitem [{\citenamefont {Choi}\ \emph {et~al.}(2015)\citenamefont {Choi},
  \citenamefont {Yao}, \citenamefont {Gopalakrishnan},\ and\ \citenamefont
  {Lukin}}]{Choi2015}%
  \BibitemOpen
  \bibfield  {author} {\bibinfo {author} {\bibfnamefont {S.}~\bibnamefont
  {Choi}}, \bibinfo {author} {\bibfnamefont {N.~Y.}\ \bibnamefont {Yao}},
  \bibinfo {author} {\bibfnamefont {S.}~\bibnamefont {Gopalakrishnan}}, \ and\
  \bibinfo {author} {\bibfnamefont {M.~D.}\ \bibnamefont {Lukin}},\ }\bibfield
  {title} {\enquote {\bibinfo {title} {Quantum control of many-body localized
  states},}\ }\href@noop {} {\bibfield  {journal} {\bibinfo  {journal} {arXiv
  preprint arXiv:1508.06992}\ } (\bibinfo {year} {2015})}\BibitemShut {NoStop}%
\bibitem [{\citenamefont {Yao}\ \emph {et~al.}(2015)\citenamefont {Yao},
  \citenamefont {Laumann},\ and\ \citenamefont {Vishwanath}}]{Yao2015}%
  \BibitemOpen
  \bibfield  {author} {\bibinfo {author} {\bibfnamefont {N.~Y.}\ \bibnamefont
  {Yao}}, \bibinfo {author} {\bibfnamefont {C.~R.}\ \bibnamefont {Laumann}}, \
  and\ \bibinfo {author} {\bibfnamefont {A.}~\bibnamefont {Vishwanath}},\
  }\bibfield  {title} {\enquote {\bibinfo {title} {Many-body localization
  protected quantum state transfer},}\ }\href@noop {} {\bibfield  {journal}
  {\bibinfo  {journal} {arXiv preprint arXiv:1508.06995}\ } (\bibinfo {year}
  {2015})}\BibitemShut {NoStop}%
\bibitem [{\citenamefont {Vasseur}\ \emph {et~al.}(2015)\citenamefont
  {Vasseur}, \citenamefont {Parameswaran},\ and\ \citenamefont
  {Moore}}]{Vasseur2015}%
  \BibitemOpen
  \bibfield  {author} {\bibinfo {author} {\bibfnamefont {R.}~\bibnamefont
  {Vasseur}}, \bibinfo {author} {\bibfnamefont {S.~A.}\ \bibnamefont
  {Parameswaran}}, \ and\ \bibinfo {author} {\bibfnamefont {J.E.}\ \bibnamefont
  {Moore}},\ }\bibfield  {title} {\enquote {\bibinfo {title} {Quantum revivals
  and many-body localization},}\ }\href@noop {} {\bibfield  {journal} {\bibinfo
   {journal} {Phys. Rev. B}\ }\textbf {\bibinfo {volume} {91}},\ \bibinfo
  {pages} {140202} (\bibinfo {year} {2015})}\BibitemShut {NoStop}%
\bibitem [{\citenamefont {Shapir}\ \emph {et~al.}(1982)\citenamefont {Shapir},
  \citenamefont {Aharony},\ and\ \citenamefont {Harris}}]{Shapir1982}%
  \BibitemOpen
  \bibfield  {author} {\bibinfo {author} {\bibfnamefont {Y.}~\bibnamefont
  {Shapir}}, \bibinfo {author} {\bibfnamefont {A.}~\bibnamefont {Aharony}}, \
  and\ \bibinfo {author} {\bibfnamefont {A~B.}\ \bibnamefont {Harris}},\
  }\bibfield  {title} {\enquote {\bibinfo {title} {Localization and quantum
  percolation},}\ }\href@noop {} {\bibfield  {journal} {\bibinfo  {journal}
  {Phys. Rev. Lett.}\ }\textbf {\bibinfo {volume} {49}},\ \bibinfo {pages}
  {486} (\bibinfo {year} {1982})}\BibitemShut {NoStop}%
\bibitem [{\citenamefont {Altland}\ and\ \citenamefont
  {Merkt}(2001)}]{Altland2001}%
  \BibitemOpen
  \bibfield  {author} {\bibinfo {author} {\bibfnamefont {A.}~\bibnamefont
  {Altland}}\ and\ \bibinfo {author} {\bibfnamefont {R.}~\bibnamefont
  {Merkt}},\ }\bibfield  {title} {\enquote {\bibinfo {title} {Spectral and
  transport properties of quantum wires with bond disorder},}\ }\href@noop {}
  {\bibfield  {journal} {\bibinfo  {journal} {Nuclear Physics B}\ }\textbf
  {\bibinfo {volume} {607}},\ \bibinfo {pages} {511--548} (\bibinfo {year}
  {2001})}\BibitemShut {NoStop}%
\bibitem [{\citenamefont {Xiong}\ and\ \citenamefont
  {Xiong}(2007)}]{Xiong2007}%
  \BibitemOpen
  \bibfield  {author} {\bibinfo {author} {\bibfnamefont {Shi-Jie}\ \bibnamefont
  {Xiong}}\ and\ \bibinfo {author} {\bibfnamefont {Ye}~\bibnamefont {Xiong}},\
  }\bibfield  {title} {\enquote {\bibinfo {title} {Anderson localization of
  electron states in graphene in different types of disorder},}\ }\href@noop {}
  {\bibfield  {journal} {\bibinfo  {journal} {Phys. Rev. B}\ }\textbf {\bibinfo
  {volume} {76}},\ \bibinfo {pages} {214204} (\bibinfo {year}
  {2007})}\BibitemShut {NoStop}%
\bibitem [{\citenamefont {Batsch}\ \emph {et~al.}(1996)\citenamefont {Batsch},
  \citenamefont {Schweitzer}, \citenamefont {Zharekeshev},\ and\ \citenamefont
  {Kramer}}]{Batsch1996}%
  \BibitemOpen
  \bibfield  {author} {\bibinfo {author} {\bibfnamefont {M.}~\bibnamefont
  {Batsch}}, \bibinfo {author} {\bibfnamefont {L.}~\bibnamefont {Schweitzer}},
  \bibinfo {author} {\bibfnamefont {I.~Kh.}\ \bibnamefont {Zharekeshev}}, \
  and\ \bibinfo {author} {\bibfnamefont {B.}~\bibnamefont {Kramer}},\
  }\bibfield  {title} {\enquote {\bibinfo {title} {Crossover from critical
  orthogonal to critical unitary statistics at the {A}nderson transition},}\
  }\href@noop {} {\bibfield  {journal} {\bibinfo  {journal} {Physical Rev.
  Lett.}\ }\textbf {\bibinfo {volume} {77}},\ \bibinfo {pages} {1552} (\bibinfo
  {year} {1996})}\BibitemShut {NoStop}%
\bibitem [{\citenamefont {Geraedts}\ \emph {et~al.}(2016)\citenamefont
  {Geraedts}, \citenamefont {Nandkishore},\ and\ \citenamefont
  {Regnault}}]{Geraedts2016}%
  \BibitemOpen
  \bibfield  {author} {\bibinfo {author} {\bibfnamefont {S.~D.}\ \bibnamefont
  {Geraedts}}, \bibinfo {author} {\bibfnamefont {R.}~\bibnamefont
  {Nandkishore}}, \ and\ \bibinfo {author} {\bibfnamefont {N.}~\bibnamefont
  {Regnault}},\ }\bibfield  {title} {\enquote {\bibinfo {title} {Many-body
  localization and thermalization: {I}nsights from the entanglement
  spectrum},}\ }\href@noop {} {\bibfield  {journal} {\bibinfo  {journal} {Phys.
  Rev. B}\ }\textbf {\bibinfo {volume} {93}},\ \bibinfo {pages} {174202}
  (\bibinfo {year} {2016})}\BibitemShut {NoStop}%
\bibitem [{\citenamefont {Dalibard}\ \emph {et~al.}(2011)\citenamefont
  {Dalibard}, \citenamefont {Gerbier}, \citenamefont {Juzeli{\=u}nas},\ and\
  \citenamefont {{\"O}hberg}}]{Dalibard2011}%
  \BibitemOpen
  \bibfield  {author} {\bibinfo {author} {\bibfnamefont {J.}~\bibnamefont
  {Dalibard}}, \bibinfo {author} {\bibfnamefont {F.}~\bibnamefont {Gerbier}},
  \bibinfo {author} {\bibfnamefont {G.}~\bibnamefont {Juzeli{\=u}nas}}, \ and\
  \bibinfo {author} {\bibfnamefont {P.}~\bibnamefont {{\"O}hberg}},\ }\bibfield
   {title} {\enquote {\bibinfo {title} {Colloquium: {A}rtificial gauge
  potentials for neutral atoms},}\ }\href@noop {} {\bibfield  {journal}
  {\bibinfo  {journal} {Rev. of Mod. Phys.}\ }\textbf {\bibinfo {volume}
  {83}},\ \bibinfo {pages} {1523} (\bibinfo {year} {2011})}\BibitemShut
  {NoStop}%
\bibitem [{\citenamefont {Goldman}\ \emph {et~al.}(2014)\citenamefont
  {Goldman}, \citenamefont {Juzeli\-{u}nas}, \citenamefont {\:Ohberg},\ and\
  \citenamefont {Spielman}}]{Goldman2014}%
  \BibitemOpen
  \bibfield  {author} {\bibinfo {author} {\bibfnamefont {N.}~\bibnamefont
  {Goldman}}, \bibinfo {author} {\bibfnamefont {G.}~\bibnamefont
  {Juzeli\-{u}nas}}, \bibinfo {author} {\bibfnamefont {P.}~\bibnamefont
  {\:Ohberg}}, \ and\ \bibinfo {author} {\bibfnamefont {I.~B.}\ \bibnamefont
  {Spielman}},\ }\bibfield  {title} {\enquote {\bibinfo {title} {Light-induced
  gauge fields for ultracold atoms},}\ }\href@noop {} {\bibfield  {journal}
  {\bibinfo  {journal} {Reports on Progress in Physics}\ }\textbf {\bibinfo
  {volume} {77}},\ \bibinfo {pages} {126401} (\bibinfo {year}
  {2014})}\BibitemShut {NoStop}%
\bibitem [{\citenamefont {Goldman}\ \emph {et~al.}(2016)\citenamefont
  {Goldman}, \citenamefont {Budich},\ and\ \citenamefont
  {Zoller}}]{Goldman2016}%
  \BibitemOpen
  \bibfield  {author} {\bibinfo {author} {\bibfnamefont {N.}~\bibnamefont
  {Goldman}}, \bibinfo {author} {\bibfnamefont {J.~C.}\ \bibnamefont {Budich}},
  \ and\ \bibinfo {author} {\bibfnamefont {P.}~\bibnamefont {Zoller}},\
  }\bibfield  {title} {\enquote {\bibinfo {title} {Topological quantum matter
  with ultracold gases in optical lattices},}\ }\href@noop {} {\bibfield
  {journal} {\bibinfo  {journal} {Nature Physics}\ }\textbf {\bibinfo {volume}
  {12}},\ \bibinfo {pages} {639--645} (\bibinfo {year} {2016})}\BibitemShut
  {NoStop}%
\bibitem [{\citenamefont {Jotzu}\ \emph {et~al.}(2014)\citenamefont {Jotzu},
  \citenamefont {Messer}, \citenamefont {Desbuquois}, \citenamefont {Lebrat},
  \citenamefont {Uehlinger}, \citenamefont {Greif},\ and\ \citenamefont
  {Esslinger}}]{Jotzu2014}%
  \BibitemOpen
  \bibfield  {author} {\bibinfo {author} {\bibfnamefont {G.}~\bibnamefont
  {Jotzu}}, \bibinfo {author} {\bibfnamefont {M.}~\bibnamefont {Messer}},
  \bibinfo {author} {\bibfnamefont {R.}~\bibnamefont {Desbuquois}}, \bibinfo
  {author} {\bibfnamefont {M.}~\bibnamefont {Lebrat}}, \bibinfo {author}
  {\bibfnamefont {T.}~\bibnamefont {Uehlinger}}, \bibinfo {author}
  {\bibfnamefont {D.}~\bibnamefont {Greif}}, \ and\ \bibinfo {author}
  {\bibfnamefont {T.}~\bibnamefont {Esslinger}},\ }\bibfield  {title} {\enquote
  {\bibinfo {title} {Experimental realization of the topological {H}aldane
  model with ultracold fermions},}\ }\href@noop {} {\bibfield  {journal}
  {\bibinfo  {journal} {Nature}\ }\textbf {\bibinfo {volume} {515}},\ \bibinfo
  {pages} {237--240} (\bibinfo {year} {2014})}\BibitemShut {NoStop}%
\bibitem [{\citenamefont {Celi}\ \emph {et~al.}(2014)\citenamefont {Celi},
  \citenamefont {Massignan}, \citenamefont {Ruseckas}, \citenamefont {Goldman},
  \citenamefont {Spielman}, \citenamefont {Juzeli{\=u}nas},\ and\ \citenamefont
  {Lewenstein}}]{Celi2014}%
  \BibitemOpen
  \bibfield  {author} {\bibinfo {author} {\bibfnamefont {A.}~\bibnamefont
  {Celi}}, \bibinfo {author} {\bibfnamefont {P.}~\bibnamefont {Massignan}},
  \bibinfo {author} {\bibfnamefont {J.}~\bibnamefont {Ruseckas}}, \bibinfo
  {author} {\bibfnamefont {N.}~\bibnamefont {Goldman}}, \bibinfo {author}
  {\bibfnamefont {I.~B.}\ \bibnamefont {Spielman}}, \bibinfo {author}
  {\bibfnamefont {G.}~\bibnamefont {Juzeli{\=u}nas}}, \ and\ \bibinfo {author}
  {\bibfnamefont {M.}~\bibnamefont {Lewenstein}},\ }\bibfield  {title}
  {\enquote {\bibinfo {title} {Synthetic gauge fields in synthetic
  dimensions},}\ }\href@noop {} {\bibfield  {journal} {\bibinfo  {journal}
  {Phys. Rev. Lett.}\ }\textbf {\bibinfo {volume} {112}},\ \bibinfo {pages}
  {043001} (\bibinfo {year} {2014})}\BibitemShut {NoStop}%
\bibitem [{Note1()}]{Note1}%
  \BibitemOpen
  \bibinfo {note} {This is to avoid the situation of being close to a point
  where momenta is a good quantum number for small disorder amplitudes in the
  case of periodic boundary conditions. That hinders the level repulsion
  characteristic of ergodic phases~\cite {Oganesyan_Huse_07}}\BibitemShut
  {NoStop}%
\bibitem [{\citenamefont {Polizzi}(2009)}]{Polizzi2009}%
  \BibitemOpen
  \bibfield  {author} {\bibinfo {author} {\bibfnamefont {E.}~\bibnamefont
  {Polizzi}},\ }\bibfield  {title} {\enquote {\bibinfo {title}
  {Density-matrix-based algorithm for solving eigenvalue problems},}\
  }\href@noop {} {\bibfield  {journal} {\bibinfo  {journal} {Phys. Rev. B}\
  }\textbf {\bibinfo {volume} {79}},\ \bibinfo {pages} {115112} (\bibinfo
  {year} {2009})}\BibitemShut {NoStop}%
\bibitem [{\citenamefont {Schnyder}\ \emph {et~al.}(2008)\citenamefont
  {Schnyder}, \citenamefont {Ryu}, \citenamefont {Furusaki},\ and\
  \citenamefont {Ludwig}}]{Schnyder2008}%
  \BibitemOpen
  \bibfield  {author} {\bibinfo {author} {\bibfnamefont {A.~P.}\ \bibnamefont
  {Schnyder}}, \bibinfo {author} {\bibfnamefont {S.}~\bibnamefont {Ryu}},
  \bibinfo {author} {\bibfnamefont {A.}~\bibnamefont {Furusaki}}, \ and\
  \bibinfo {author} {\bibfnamefont {A.~W.W.}\ \bibnamefont {Ludwig}},\
  }\bibfield  {title} {\enquote {\bibinfo {title} {Classification of
  topological insulators and superconductors in three spatial dimensions},}\
  }\href@noop {} {\bibfield  {journal} {\bibinfo  {journal} {Phys. Rev. B}\
  }\textbf {\bibinfo {volume} {78}},\ \bibinfo {pages} {195125} (\bibinfo
  {year} {2008})}\BibitemShut {NoStop}%
\bibitem [{\citenamefont {Iyer}\ \emph {et~al.}(2013)\citenamefont {Iyer},
  \citenamefont {Oganesyan}, \citenamefont {Refael},\ and\ \citenamefont
  {Huse}}]{Iyer2013}%
  \BibitemOpen
  \bibfield  {author} {\bibinfo {author} {\bibfnamefont {Shankar}\ \bibnamefont
  {Iyer}}, \bibinfo {author} {\bibfnamefont {Vadim}\ \bibnamefont {Oganesyan}},
  \bibinfo {author} {\bibfnamefont {Gil}\ \bibnamefont {Refael}}, \ and\
  \bibinfo {author} {\bibfnamefont {David~A.}\ \bibnamefont {Huse}},\
  }\bibfield  {title} {\enquote {\bibinfo {title} {Many-body localization in a
  quasiperiodic system},}\ }\href {\doibase 10.1103/PhysRevB.87.134202}
  {\bibfield  {journal} {\bibinfo  {journal} {Phys. Rev. B}\ }\textbf {\bibinfo
  {volume} {87}},\ \bibinfo {pages} {134202} (\bibinfo {year}
  {2013})}\BibitemShut {NoStop}%
\bibitem [{\citenamefont {Tang}\ \emph {et~al.}(2015)\citenamefont {Tang},
  \citenamefont {Iyer},\ and\ \citenamefont {Rigol}}]{Baoming2015}%
  \BibitemOpen
  \bibfield  {author} {\bibinfo {author} {\bibfnamefont {Baoming}\ \bibnamefont
  {Tang}}, \bibinfo {author} {\bibfnamefont {Deepak}\ \bibnamefont {Iyer}}, \
  and\ \bibinfo {author} {\bibfnamefont {Marcos}\ \bibnamefont {Rigol}},\
  }\bibfield  {title} {\enquote {\bibinfo {title} {Quantum quenches and
  many-body localization in the thermodynamic limit},}\ }\href {\doibase
  10.1103/PhysRevB.91.161109} {\bibfield  {journal} {\bibinfo  {journal} {Phys.
  Rev. B}\ }\textbf {\bibinfo {volume} {91}},\ \bibinfo {pages} {161109}
  (\bibinfo {year} {2015})}\BibitemShut {NoStop}%
\bibitem [{\citenamefont {Serbyn}\ and\ \citenamefont
  {Moore}(2016)}]{Serbym2016}%
  \BibitemOpen
  \bibfield  {author} {\bibinfo {author} {\bibfnamefont {Maksym}\ \bibnamefont
  {Serbyn}}\ and\ \bibinfo {author} {\bibfnamefont {Joel~E.}\ \bibnamefont
  {Moore}},\ }\bibfield  {title} {\enquote {\bibinfo {title} {Spectral
  statistics across the many-body localization transition},}\ }\href {\doibase
  10.1103/PhysRevB.93.041424} {\bibfield  {journal} {\bibinfo  {journal} {Phys.
  Rev. B}\ }\textbf {\bibinfo {volume} {93}},\ \bibinfo {pages} {041424}
  (\bibinfo {year} {2016})}\BibitemShut {NoStop}%
\bibitem [{\citenamefont {Atas}\ \emph {et~al.}(2013)\citenamefont {Atas},
  \citenamefont {Bogomolny}, \citenamefont {Giraud},\ and\ \citenamefont
  {Roux}}]{Atas2013}%
  \BibitemOpen
  \bibfield  {author} {\bibinfo {author} {\bibfnamefont {Y.~Y.}\ \bibnamefont
  {Atas}}, \bibinfo {author} {\bibfnamefont {E.}~\bibnamefont {Bogomolny}},
  \bibinfo {author} {\bibfnamefont {O.}~\bibnamefont {Giraud}}, \ and\ \bibinfo
  {author} {\bibfnamefont {G.}~\bibnamefont {Roux}},\ }\bibfield  {title}
  {\enquote {\bibinfo {title} {Distribution of the ratio of consecutive level
  spacings in random matrix ensembles},}\ }\href@noop {} {\bibfield  {journal}
  {\bibinfo  {journal} {Phys. Rev. Lett.}\ }\textbf {\bibinfo {volume} {110}},\
  \bibinfo {pages} {084101} (\bibinfo {year} {2013})}\BibitemShut {NoStop}%
\bibitem [{\citenamefont {Nandkishore}\ and\ \citenamefont
  {Potter}(2014)}]{Nandkishore2014}%
  \BibitemOpen
  \bibfield  {author} {\bibinfo {author} {\bibfnamefont {R.}~\bibnamefont
  {Nandkishore}}\ and\ \bibinfo {author} {\bibfnamefont {A.~C.}\ \bibnamefont
  {Potter}},\ }\bibfield  {title} {\enquote {\bibinfo {title} {Marginal
  anderson localization and many-body delocalization},}\ }\href@noop {}
  {\bibfield  {journal} {\bibinfo  {journal} {Phys. Rev. B}\ }\textbf {\bibinfo
  {volume} {90}},\ \bibinfo {pages} {195115} (\bibinfo {year}
  {2014})}\BibitemShut {NoStop}%
\bibitem [{\citenamefont {Altland}\ and\ \citenamefont
  {Simons}(1999)}]{Altland1999}%
  \BibitemOpen
  \bibfield  {author} {\bibinfo {author} {\bibfnamefont {A.}~\bibnamefont
  {Altland}}\ and\ \bibinfo {author} {\bibfnamefont {B.~D.}\ \bibnamefont
  {Simons}},\ }\bibfield  {title} {\enquote {\bibinfo {title} {Field theory of
  the random flux model},}\ }\href
  {http://stacks.iop.org/0305-4470/32/i=31/a=101} {\bibfield  {journal}
  {\bibinfo  {journal} {Nucl. Phys. B}\ }\textbf {\bibinfo {volume} {562}},\
  \bibinfo {pages} {445} (\bibinfo {year} {1999})},\ \bibinfo {note} {and
  references therein}\BibitemShut {NoStop}%
\bibitem [{Note2()}]{Note2}%
  \BibitemOpen
  \bibinfo {note} {Typical equilibration times are defined as when the smallest
  time in which results of the relaxation dynamics remains closer to the DE
  prediction.}\BibitemShut {Stop}%
\bibitem [{\citenamefont {Serbyn}\ \emph
  {et~al.}(2014{\natexlab{b}})\citenamefont {Serbyn}, \citenamefont
  {Papi\ifmmode~\acute{c}\else \'{c}\fi{}},\ and\ \citenamefont
  {Abanin}}]{Serbym2014b}%
  \BibitemOpen
  \bibfield  {author} {\bibinfo {author} {\bibfnamefont {Maksym}\ \bibnamefont
  {Serbyn}}, \bibinfo {author} {\bibfnamefont {Z.}~\bibnamefont
  {Papi\ifmmode~\acute{c}\else \'{c}\fi{}}}, \ and\ \bibinfo {author}
  {\bibfnamefont {D.~A.}\ \bibnamefont {Abanin}},\ }\bibfield  {title}
  {\enquote {\bibinfo {title} {Quantum quenches in the many-body localized
  phase},}\ }\href {\doibase 10.1103/PhysRevB.90.174302} {\bibfield  {journal}
  {\bibinfo  {journal} {Phys. Rev. B}\ }\textbf {\bibinfo {volume} {90}},\
  \bibinfo {pages} {174302} (\bibinfo {year} {2014}{\natexlab{b}})}\BibitemShut
  {NoStop}%
\bibitem [{Note3()}]{Note3}%
  \BibitemOpen
  \bibinfo {note} {The Hamiltonian in the present form still possess further
  symmetries which are removed by inserting small symmetry breaking fields (see
  Appendix B).}\BibitemShut {Stop}%
\bibitem [{\citenamefont {Rigol}(2009)}]{rigol_09a}%
  \BibitemOpen
  \bibfield  {author} {\bibinfo {author} {\bibfnamefont {M.}~\bibnamefont
  {Rigol}},\ }\bibfield  {title} {\enquote {\bibinfo {title} {Breakdown of
  thermalization in finite one-dimensional systems},}\ }\href {\doibase
  10.1103/PhysRevLett.103.100403} {\bibfield  {journal} {\bibinfo  {journal}
  {Phys. Rev. Lett.}\ }\textbf {\bibinfo {volume} {103}},\ \bibinfo {pages}
  {100403} (\bibinfo {year} {2009})}\BibitemShut {NoStop}%
\bibitem [{\citenamefont {Bloch}\ \emph {et~al.}(2008)\citenamefont {Bloch},
  \citenamefont {Dalibard},\ and\ \citenamefont {Zwerger}}]{Bloch2008}%
  \BibitemOpen
  \bibfield  {author} {\bibinfo {author} {\bibfnamefont {I.}~\bibnamefont
  {Bloch}}, \bibinfo {author} {\bibfnamefont {J.}~\bibnamefont {Dalibard}}, \
  and\ \bibinfo {author} {\bibfnamefont {W.}~\bibnamefont {Zwerger}},\
  }\bibfield  {title} {\enquote {\bibinfo {title} {Many-body physics with
  ultracold gases},}\ }\href@noop {} {\bibfield  {journal} {\bibinfo  {journal}
  {Reviews of Modern Physics}\ }\textbf {\bibinfo {volume} {80}},\ \bibinfo
  {pages} {885} (\bibinfo {year} {2008})}\BibitemShut {NoStop}%
\bibitem [{\citenamefont {Beugeling}\ \emph {et~al.}(2014)\citenamefont
  {Beugeling}, \citenamefont {Moessner},\ and\ \citenamefont
  {Haque}}]{beugeling_moessner_14}%
  \BibitemOpen
  \bibfield  {author} {\bibinfo {author} {\bibfnamefont {W.}~\bibnamefont
  {Beugeling}}, \bibinfo {author} {\bibfnamefont {R.}~\bibnamefont {Moessner}},
  \ and\ \bibinfo {author} {\bibfnamefont {M.}~\bibnamefont {Haque}},\
  }\bibfield  {title} {\enquote {\bibinfo {title} {Finite-size scaling of
  eigenstate thermalization},}\ }\href {\doibase 10.1103/PhysRevE.89.042112}
  {\bibfield  {journal} {\bibinfo  {journal} {Phys. Rev. E}\ }\textbf {\bibinfo
  {volume} {89}},\ \bibinfo {pages} {042112} (\bibinfo {year}
  {2014})}\BibitemShut {NoStop}%
\bibitem [{\citenamefont {Kim}\ \emph {et~al.}(2014)\citenamefont {Kim},
  \citenamefont {Ikeda},\ and\ \citenamefont {Huse}}]{Kim2014}%
  \BibitemOpen
  \bibfield  {author} {\bibinfo {author} {\bibfnamefont {H.}~\bibnamefont
  {Kim}}, \bibinfo {author} {\bibfnamefont {T.~N.}\ \bibnamefont {Ikeda}}, \
  and\ \bibinfo {author} {\bibfnamefont {D.~A.}\ \bibnamefont {Huse}},\
  }\bibfield  {title} {\enquote {\bibinfo {title} {Testing whether all
  eigenstates obey the eigenstate thermalization hypothesis},}\ }\href@noop {}
  {\bibfield  {journal} {\bibinfo  {journal} {Phys. Rev. E}\ }\textbf {\bibinfo
  {volume} {90}},\ \bibinfo {pages} {052105} (\bibinfo {year}
  {2014})}\BibitemShut {NoStop}%
\bibitem [{\citenamefont {Mondaini}\ \emph {et~al.}(2016)\citenamefont
  {Mondaini}, \citenamefont {Fratus}, \citenamefont {Srednicki},\ and\
  \citenamefont {Rigol}}]{Mondaini2016}%
  \BibitemOpen
  \bibfield  {author} {\bibinfo {author} {\bibfnamefont {R.}~\bibnamefont
  {Mondaini}}, \bibinfo {author} {\bibfnamefont {K.~R.}\ \bibnamefont
  {Fratus}}, \bibinfo {author} {\bibfnamefont {M.}~\bibnamefont {Srednicki}}, \
  and\ \bibinfo {author} {\bibfnamefont {M.}~\bibnamefont {Rigol}},\ }\bibfield
   {title} {\enquote {\bibinfo {title} {Eigenstate thermalization in the
  two-dimensional transverse field {I}sing model},}\ }\href@noop {} {\bibfield
  {journal} {\bibinfo  {journal} {Phys. Rev. E}\ }\textbf {\bibinfo {volume}
  {93}},\ \bibinfo {pages} {032104} (\bibinfo {year} {2016})}\BibitemShut
  {NoStop}%
\end{thebibliography}%

\end{document}